\documentclass[twocolumn]{aastex62}
\usepackage{amsmath}
\usepackage{amsfonts}
\usepackage{amssymb}
\usepackage{ulem}
\usepackage{multirow}

\newcommand{\torb}{{t_{\rm orb}}}
\newcommand{\Kpr}{{K^\prime}}

\received{XXX}
\revised{XXX}
\accepted{XXX}

\shorttitle{Gap Parameter Relations}
\shortauthors{Yun et al.}

\begin{document}

\title{Properties of Density and Velocity Gaps Induced by a Planet in a Protoplanetary Disk}



\author[0000-0003-4353-294X]{Han Gyeol Yun}
\affiliation{Department of Physics \& Astronomy, Seoul National University, Seoul 08826, Korea}
\author[0000-0003-4625-229X]{Woong-Tae Kim}
\affiliation{Department of Physics \& Astronomy, Seoul National University, Seoul 08826, Korea}
\affiliation{Center for Theoretical Physics (CTP), Seoul National University, Seoul 08826, Korea}
\author[0000-0001-7258-770X]{Jaehan Bae}
\affiliation{Department of Terrestrial Magnetism, Carnegie Institution for Science, 5241 Broad Branch Road NW, Washington, DC 20015, USA}
\author[0000-0002-2641-9964]{Cheongho Han}
\affiliation{Department of Physics, Chungbuk National University, Cheongju 28644, Korea}

\email{hangyeol@snu.ac.kr, wkim@astro.snu.ac.kr}
\email{jbae@carnegiescience.edu, cheongho@astroph.chungbuk.ac.kr,}

\begin{abstract}
Gravitational interactions between a protoplanetary disk and its embedded planet is one of the formation mechanisms of gaps and rings found in recent ALMA observations. To quantify the gap properties measured in not only surface density but also rotational velocity profiles, we run two-dimensional hydrodynamic simulations of protoplanetary disks by varying three parameters: the mass ratio $q$ of a planet to a central star, the ratio of the disk scale height $h_p$ to the orbital radius $r_p$ of the planet, and the viscosity parameter $\alpha$.  We find the gap depth $\delta_\Sigma$ in the gas surface density depends on a single dimensionless parameter $K\equiv q^2(h_p/r_p)^{-5}\alpha^{-1}$ as
$\delta_\Sigma = (1 + 0.046K)^{-1}$, consistent with the previous results of \citet{ka2015a}. The gap depth $\delta_V$ in the rotational velocity is given by $\delta_V= 0.007 (h_p/r_p) K^{1.38}/(1 +0.06K^{1.03})$. The gap width, in both surface density and rotational velocity, has a minimum of about $4.7 h_p$ when the planet mass $M_p$ is around the disk thermal mass $M_{\text{th}}$, while it increases in a power-law fashion as $M_p/M_{\text{th}}$ increases or decrease from unity. Such a minimum in the gap width arises because spirals from sub-thermal planets have to propagate before they shock the disk gas and open a gap. We compare our relations for the gap depth and width with the previous results, and discuss their applicability to observations.
\end{abstract}


\section{Introduction} \label{sec:intro}

High resolution observations of protoplanetary disks in the past decade have found diverse substructures in the disks, including spiral arms (e.g., SAO 206462, \citealp{mu2012}; MWC758, \citealp{gr2013}; HD 100453, \citealp{wa2015}), large-scale asymmetries (e.g., HD 142527, \citealp{ca2013}; Oph IRS 48, \citealp{van2013}), and gaps or rings (e.g., HL Tau, \citealp{al2015}, TW Hya: \citealp{an2016}; HD 163296, \citealp{is2016}; HD 169142, \citealp{fe2017}; AA Tau, \citealp{lo2017}; Elias 2-24, \citealp{ci2017}; AS 209, \citealp{fe2018}; GY 91, \citealp{se2018}; V1094 Scorpii, \citealp{va2018}; HD 143005, \citealp{be2018}; HD 92945, \citealp{ma2019}). Compared to other substructures in the disks, gaps or rings are nearly axisymmetric and concentric.

While these substructures appear common in protoplanetary disks, their physical origin has remained uncertain. For gaps or rings, in particular, a number of scenarios have been proposed as their formation mechanisms. For example, fast pebble growth near the snowlines of abundant volatile molecules was proposed by \citet {zh2015} to explain the observed gaps in HL Tau. However, the presence of eccentric rings \citep{do2018a} and recent observations that gap locations do not correspond to the snowlines of the most common species in many proptoplanetary disks \citep{lo2018,hu2018,van2019} suggest that the snowline scenario is unlikely as a common origin of multiple gaps. Other potential mechanisms include secular gravitational instability \citep{ti2016}, toroidal vortices induced by large-scale instability \citep{lb2016}, self-induced dust vortex \citep{go2015}, disk winds \citep{su2017}, zonal flow \citep{fl2015}, and sintering-induced piling-up of dust aggregates \citep{ok2016}.  Although these mechanisms successfully produce rings at the locations close the observed ring radii in specific systems, it is unclear whether they can be applicable to all observed protoplanetary disks with gaps \citep{hu2018}.

Perhaps, the most natural and favored mechanism for gaps/rings may be gravitational interactions between the disk and its embedded planet(s) \citep{lp1979}. Density wakes launched by the gravity of the planet can transfer angular momentum from the regions inside the planet orbit to the outside, making the gas in the disk pushed away from the vicinity of the planet \citep{gt1980,ra2002}. In fact, \citet{ba2018} showed that the disk-planet interactions with differing parameters such as viscosity and dust distribution, etc.\ can create diverse morphology that includes a full disk, a transition disk with an inner cavity, a disk with a single gap and a central continuum peak and a disk with multiple gap and a central continuum peak. Even a single planet can forms multiple gaps in a low-viscosity disk \citep{do2017,ba2017}, because secondary and tertiary spiral arms can also grow enough to induce shocks across which gas loses its angular momentum (see also \citealp{bz2018,mr2019}).

The shape of a gap produced by a planet is determined by the balance between the tidal torque density and the viscous stress. In a disk with scale height $h_p$ and surface density $\Sigma$ around a protostar with mass $M_*$, the tidal torque by a planet with mass $M_p$ at orbital radius $r_p$ is proportional to $q^2 (h_p/r_p)^{-3}\Sigma$ with $q\equiv M_p/M_*$ (e.g., \citealt{gt1980,pap84}), while the viscous stress is proportional to the viscous parameters $\alpha$ of \citet{ss1973}. It was shown that the gap depth in the surface density can be characterized by a single dimensionless parameter $K\equiv q^2(h_p/r_p)^{-5}\alpha^{-1}$ \citep{dm2013,fu2014,ka2015a}, while the gap width can be described solely by $K'\equiv K(h_p/r_p)^2$ \citep{ka2016,ka2017}. These relations were applied to constrain the masses of embedded planets in several systems such as HL Tau \citep{ka2015b,ka2016}, HD169142 \citep{ka2015b}, and HD 97048 \citep{gi2016}.

However, applying these relations for the gap depth and width to observations requires a strict assumption that the distribution of dust particles is well-matched with that of gas \citep{ka2015b}. In reality, the conversion of observed dust continuum to the gas surface density is subject to many uncertainties surrounding dust-to-gas ratio, varying dust properties, and chemical effects \citep{be2013,mi2017}. To directly measure the gap properties in dust continuum profiles, \citet{zha2018} ran numerical simulations by including dust particles and obtained the empirical relations for the gap depth and width in terms of the various dimensionless parameters.
But, they were still unable to incorporate dust evolution, feedback to the gas, and the potential effects of streaming instability in the simulations
that may affect the gap properties significantly.

One way to circumvent the uncertainties in the gap parameters measured from the gas surface density profiles is to use the rotational velocity obtained from gas tracers such as CO that directly probes the kinematic changes in the gas disk induced by an embedded planet. \citet{pe2015,pe2018} showed that kinematic features including circumplanetary disk and large-scale velocity perturbations induced by a Jupiter-mass planet are observable with the ALMA.
In fact, \citet{te2018} and \citet{ke2019} recently compared the observed rotational velocity profiles with the numerical simulations to infer the masses of planets in HD 163296 and PDS 70, respectively. \citet{zha2018} ran extensive numerical simulations to find an empirical relation for the amplitude of the perturbed rotational velocity as a combination of $q$, $h_p/r_p$ and $\alpha$. Since the simulations were run up to $10^3$ planetary orbits, however, it is uncertain whether the gaps in their models reach a steady state, as they noted (see also \citealt{ro2016}). In addition, they found that the width of velocity gaps is roughly 4.4 times $h_p$, insensitive to $q$ and $\alpha$, which needs to be checked in long-term evolution.

In this paper, we run hydrodynamic simulations of protoplanetary disks to systematically investigate the gap properties in not only gas surface density profile but also rotational velocity profile induced by an embedded planet. We vary three parameters, $q$, $h_p/r_p$, and $\alpha$, in a wide range, and explore how the gap depth and width depend on these parameters. Our work extends \citet{zha2018} by exploring a wider range of the parameter space and by running the simulations 10 times longer than their models in order to achieve quasi-steady configurations of the gaps. We also introduce a new definition of the gap width in the surface density and rotational velocity profiles and provide the physical explanation for its dependence on the planet mass.

The rest of this paper is organized as follows. In Section \ref{sec:Method}, we describe our simulation setups and model parameters. In Section \ref{sec:Results}, we present the gap properties in the surface density rotation velocity profiles. We discuss  our results in Sections \ref{sec:Discussion} and give our conclusions in \ref{sec:Summary}.

\section{Numerical Method} \label{sec:Method}

We consider a protoplanetary disk rotating at angular frequency $\Omega$ about a central protostar with mass $M_*$. The disk is assumed to be razor-thin along the vertical direction, unmagnetized, and non-self-gravitating. To
study gravitational interactions between the disk with an embedded planet with mass $M_p$, we run two-dimensional (2D) hydrodynamic simulations using FARGO3D in cylindrical polar coordinates $(r, \phi)$ \citep{m2000,bm2016}. We do not consider the effects of dust and planet migration in the present work. The basic equations we solve are
\begin{gather}
\frac{\partial\Sigma}{\partial t}+\nabla\cdot (\Sigma \mathbf{v}) =0, \label{eq:con}\\
\left(\frac{\partial \mathbf{v}}{\partial t}+\mathbf{v}\cdot\nabla \mathbf{v}\right)=- \frac{1}{\Sigma} \nabla P -\nabla(\Phi_{*}+\Phi_p)-\frac{1}{\Sigma}\nabla\cdot\boldsymbol{\Pi},
\label{eq:mom}
\end{gather}
where $\Sigma$ is the surface density, $\mathbf{v}$ is the velocity, and $P \equiv c_s^2\Sigma$ is a vertically integrated gas pressure with $c_s$ being the isothermal speed of sound. The pressure scale height of the disk is given by $h=c_s/\Omega_K$, where $\Omega_K=(GM_*/r^3)^{1/2}$ is the angular velocity of Keplerian rotation. In Equation \eqref{eq:mom}, $\Phi_*$ and $\Phi_p$ are the gravitational potentials of the central star and the planet located at $\mathbf{r}=\mathbf{r}_p$, respectively, given by
\begin{gather}
\Phi_*=-\frac{GM_*}{|\mathbf{r}|}\quad \text{and}\quad
\Phi_p=-\frac{GM_p}{\sqrt{|\mathbf{r}-\mathbf{r}_p|^2+s^2}},
\end{gather}
where $s$ is the softening length taken equal to $0.6h_p$ for $h_p\equiv h(r_p)$. The planet is set to follow the Keplerian rotation with angular velocity $\Omega_{p}\equiv\Omega_K(r_p)$, without undergoing migration.
We ignore the indirect term arising from the motions of the central star relative to the center of mass of the whole system, which are shown to make insignificant differences on the gap properties (Appendix \ref{ap:indirect}; see also \citealp{ka2017})

The last term in Equation \eqref{eq:mom} represents the viscous stress tensor
\begin{equation}
\boldsymbol{\Pi}=\nu\Sigma\left[\nabla\mathbf{v}+(\nabla\mathbf{v})^{\text{T}}-\frac{2}{3}(\nabla\cdot\mathbf{v})\pmb{\mathbb{I}}\right],
\end{equation}
where $\nu$ is the kinematic viscosity and $\pmb{\mathbb{I}}$ is the identity matrix. We adopt an $\alpha$-disk model of \citet{ss1973} with  $\nu = \alpha c_s^2/\Omega$, and vary $\alpha$ to control the strength of the viscosity.

The density distribution of our initial disk follows a power-law with an exponential cutoff:
\begin{equation}\label{eq:sigmao}
\Sigma_0(r)=\Sigma_0(r_p)\left(\frac{r}{r_p}\right)^{-m}
\exp\left[1-\left(\frac{r}{r_p}\right)^{2-m}\right],
\end{equation}
corresponding to a quasi-equilibrium solution of viscous disks (e.g., \citealt{lp1974}).
The temperature profile $T(r)$ is set to a simple power-law
\begin{equation}
T_0(r)=T_0(r_p)\left(\frac{r}{r_p}\right)^{-n},
\label{eq:eqT}
\end{equation}
which remains unchanged over time in our simulations.
In this paper, we adopt $m=1$ and $n=0.5$.
These radial density and temperature distributions describe the observed protoplanetary disks reasonably well (e.g., \citealt{an2009,an2010}).
We vary $T_0(r_p)$ or $h_p$ to explore disks with differing temperature. In what follows, the non-uniform disks refer to a power-law
disk with an exponential cutoff, in contrast to uniform disks with constant density and temperature (e.g., \citealt{ka2015a,ka2016,ka2017}).

The initial rotational velocity $v_{\phi,0}$ of gas in equilibrium is very close to the Keplerian velocity (within $\sim 6\%$ of $r\Omega_K$).
Our simulation domain extends from $r=0.3r_p$ to $r=3r_p$ in radius and from 0 to $2\pi$ in azimuth. For the boundary conditions, we adopt the wave-damping zones at $0.3r_p\leq r\leq 0.36r_p$ and $2.7r_p\leq r \leq 3.0r_p $, which is known to prevent wave reflections at the boundaries \citep{de2006}.
For simulations presented in this paper, we set up a non-uniform, logarithmically spaced cylindrical grid with $N_r=512$ radial zones and $N_\phi=$1396 azimuthal zones. This makes the zones almost square-shaped throughout the grid (i.e., $r\Delta\phi/\Delta r\approx 1$). The grid spacing adopted here results from a compromise between computational cost and accuracy. By running simulations with various resolution, we checked that the results with $N_r=512$ agrees with those $N_r=1024$ within $\sim 4\%$.

The fundamental dimensional units for length, time, and mass are the orbital radius $r_p$ and orbital time $t_\text{orb}=2\pi/\Omega_{p}$ of the planet, and the mass of the central star $M_*$. Then, Equations \eqref{eq:con} and \eqref{eq:mom} in dimensionless form depend only on three dimensionless parameters: the mass ratio $q\equiv M_p/M_*$, the disk aspect ratio $h_p/r_p$, and the viscosity parameter $\alpha$. We run a total of 72 simulations that differ in these three parameters. The planet mass is varied in the range between $3\times10^{-5}$ and $3\times10^{-3}$ relative to $M_*$, or between $0.3$ and  $9.0$ relative to the thermal mass $M\textsubscript{th}\equiv M_*(h_p/r_p)^{3}$ (e.g., \citealt{gr2001}). We take $0.03, 0.05$, $0.07, 0.10, 0.12$ for $h_p/r_p$, and $3\times10^{-4}, 6\times10^{-4}, 1\times10^{-3}, 3\times10^{-3}$ for the $\alpha$ parameter. All simulations are run up to $t=(10^4+10^2)\torb$. Table \ref{tb:param} in Appendix \ref{ap:tbl} lists the model parameters and the measured gap properties.

\section{Simulation Results} \label{sec:Results}

\begin{figure*}
	\centering
	\includegraphics[width=1.0\linewidth]{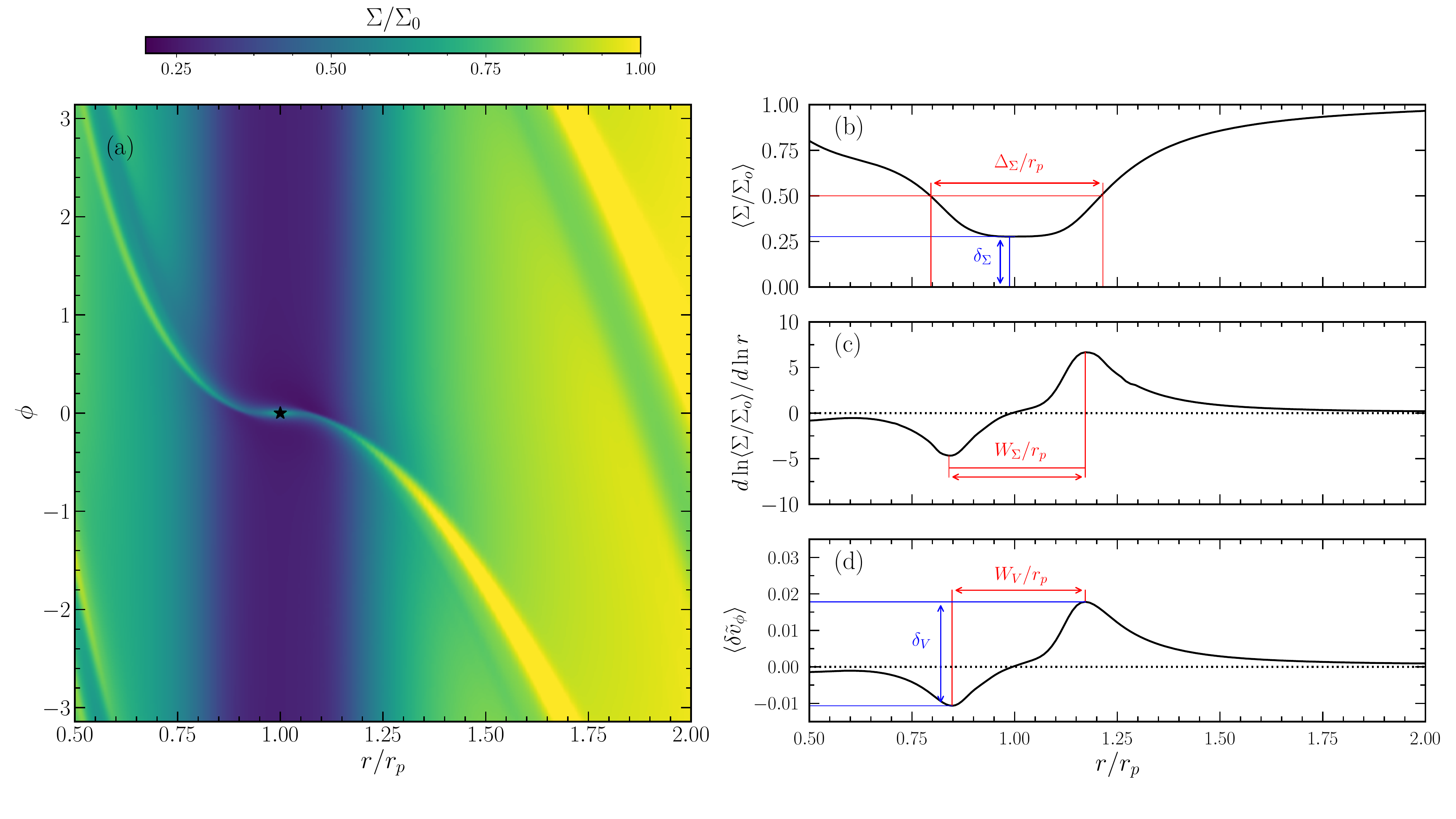}
	\vspace*{-7.5mm}
	\caption{Distributions of the perturbed density and azimuthal velocity for a model with $q=3\times10^{-4}$ (or $M_p/M_\text{th}=0.87$), $h_p/r_p=0.07$, and $\alpha=1\times 10^{-3}$. (a) Time-averaged distribution of $\Sigma/\Sigma_0$ in the $r$--$\phi$ plane, with the planet location is marked by a dark star symbol at $r=r_p$ and $\phi=0$.  (b) Radial distribution of the azimuthally averaged quantity $\langle\Sigma/\Sigma_0\rangle$, with $\Delta_\Sigma$ and $\delta_\Sigma$ illustrating the definitions of \citet{ka2015a,ka2016} for the gap width and depth, respectively. (c) Radiation distribution of the logarithmic gradient of $\langle\Sigma/\Sigma_0\rangle$. The new gap width $W_\Sigma$ defined as the distance between the extrema of $d\ln\langle\Sigma/\Sigma_0\rangle/d\ln r$ is indicated. (d) Radial distribution of the perturbed rotational velocity. The dimensionless amplitude $\delta_V$ of the perturbed velocity and the width $W_V$ of the perturbed regions are indicated.}
	\label{fig:Surfsnap}
\end{figure*}

An introduction of the gravitational potential of a planet excites spiral waves in the disk that eventually develop into shocks. For a low-mass planet, perturbations are weak so that they propagate radially away from the planet before turning to shocks \citep{gr2001}. When a planet is massive, however, the shock formation occurs almost instantly near the planet location. Almost inviscid disks with small $\alpha \,(\lesssim 10^{-4})$ may produce up to three spiral shocks, while viscious disks with large $\alpha$ considered here form only one spiral shock \citep{ba2017}. When the gas inside (outside) the orbit of the planet experiences a spiral shock, it loses (gains) angular momentum and thus moves inward (outward) in the radial direction, producing a gap in the surface density profile \citep{ra2002}. Similarly, the disk rotation curve, which is initially close to Keplerian, is also perturbed to become sub- and super-Keplerian in the regions with $r<r_p$ and $r>r_p$, respectively.

The disk reaches a quasi-steady equilibrium by $t\sim10^4t_\text{orb}$ (see Appendix \ref{ap:time}).  To quantify the gap properties, we select 11 snapshots from $t=10^4\torb$ to $t=(10^4+10^2)\torb$, separated by a time interval $\Delta t=10\torb$, and take their time averages.  We then remove the disk material, within the distance $d=2 \max\left[h_p,(M_p/3M_*)^{1/3}\right]$ from the planet, that belongs to the spiral shocks attached to the planet rather than the gap \citep{fu2014}. Figure \ref{fig:Surfsnap} plots the time-averaged distribution of the normalized surface density $\Sigma/\Sigma_0$ in the $r$--$\phi$ plane as well as the radial distributions of $\langle\Sigma/\Sigma_0\rangle$, $d\ln\langle\Sigma/\Sigma_0\rangle/d\ln r$, and $\langle\delta \tilde{v}_{\phi}\rangle = \langle (v_{\phi}-v_{\phi,0})/v_{\phi,0} \rangle$ for a model with $q=3\times 10^{-4}$ (or $M_p/M_\text{th}=0.87$), $h_p/r_p=0.07$, and $\alpha=1\times 10^{-3}$. Here, the angle brackets $\langle \,\rangle$ denote the temporal and azimuthal average. In what follows, we first present the dependence on the input parameters of the gap depth and width in the $\langle \Sigma/\Sigma_0\rangle$ distributions. We then discuss the depth and width in the perturbed velocity profiles.

\subsection{Gap in Surface Density} \label{ssec:surf}

Here we focus on the gap depth ($\delta_{\Sigma}$) and the width ($\Delta_{\Sigma}$) in the surface density profiles and explore their dependence on the combinations of the dimensionless parameters $q$, $h_p/r_p$, and $\alpha$.

\subsubsection{Gap Depth} \label{sssec:sdd}

\citet{ka2015a} defined the gap depth in the surface density as $\delta_\Sigma\equiv \min \langle\Sigma/\Sigma_0\rangle$, as illustrated in Figure \ref{fig:Surfsnap}(b).
For disks with uniform density and temperature distributions, they showed that $\delta_{\Sigma}$ depends on $q$, $h_p/r_p$, and $\alpha$ through a single dimensionless parameter $K\equiv q^2(h_p/r_p)^{-5}\alpha^{-1}$.
From the requirement that the planet-induced gravitational torque balances the viscous torque in the linear analysis, \citet{ka2015a} derived the relation
\begin{equation}\label{eq:Surf_depth_K15}
\delta_{\Sigma}^{\rm K15} \approx\frac{1}{1+0.040K},
\end{equation}
consistent with the results of their numerical simulations for uniform disks.

\begin{figure}
	\centering
	\includegraphics[width=1.0\linewidth]{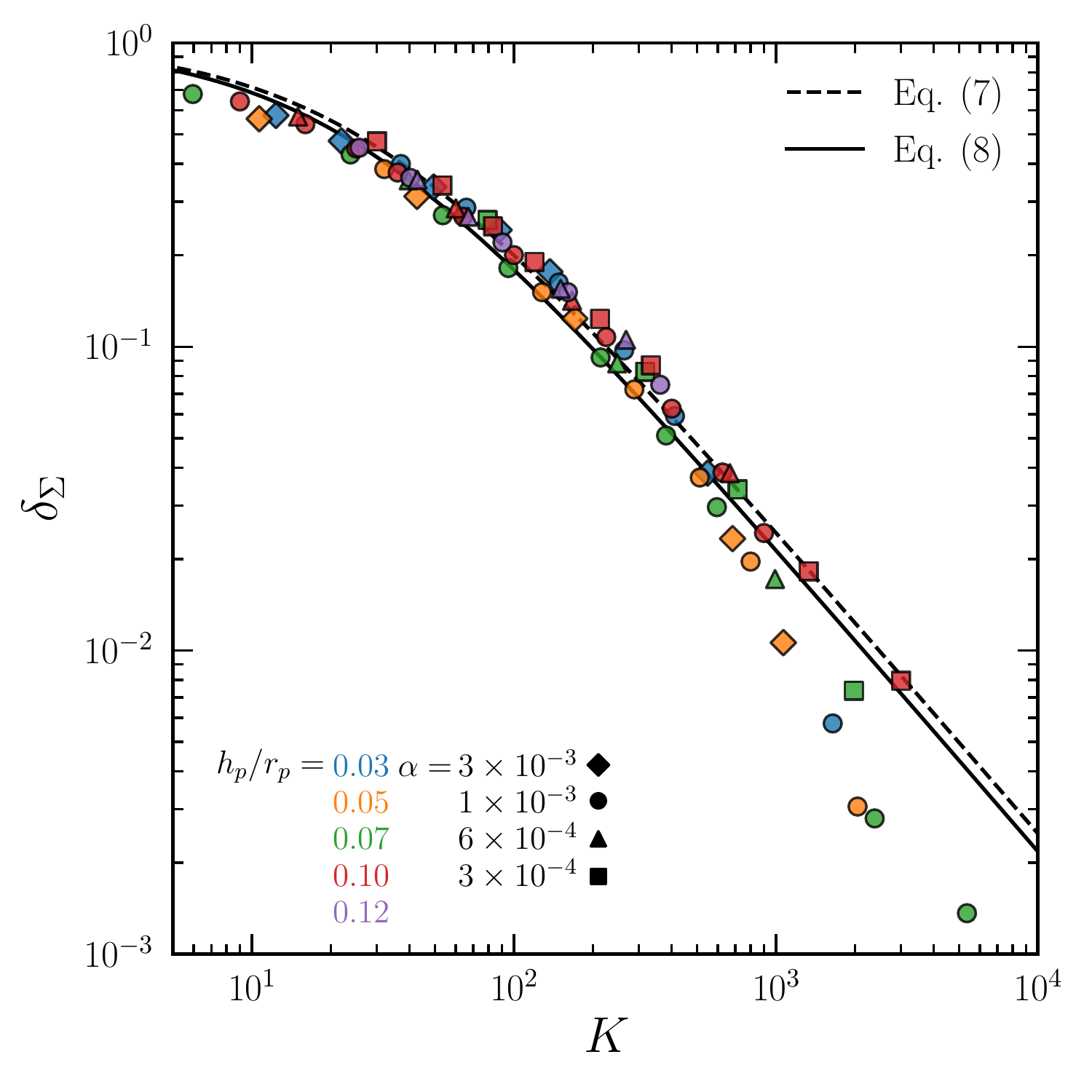}
	\caption{Gap depth $\delta_{\Sigma}$, based on the definition of \citet{ka2015a}, measured from our simulations as a function of $K\equiv q^2(h_p/r_p)^{-5}\alpha^{-1}$. The solid line is our fit (Equation \eqref{eq:Surf_depth}) for $K<10^3$ to the numerical results for non-uniform disks, while the dotted line draws Equation \eqref{eq:Surf_depth_K15} for uniform disks.}
	\label{fig:smin}
\end{figure}

To explore how the gap depth depend on $K$ in our non-uniform disks, Figure \ref{fig:smin} plots the measured $\delta_{\Sigma}$ as a function of $K$ for all models. We fit the data using a functional form same as in Equation \eqref{eq:Surf_depth_K15} but with a different coefficient. The solid line draws our least-square fit
\begin{equation}\label{eq:Surf_depth}
\delta_{\Sigma}=\frac{1}{1+0.046K},
\end{equation}
to the numerical results for $K<10^3$.\footnote{Note that  $\delta_\Sigma$ actually measures the height of a density floor from the bottom in the $\langle \Sigma/\Sigma_0\rangle$ distribution. The \emph{real} gap depth relative to the unperturbed value is $1-\delta_{\Sigma} = 0.046K/(1+0.046K)$.}  Equation \eqref{eq:Surf_depth} is almost equal to Equation \eqref{eq:Surf_depth_K15}, shown as the dotted line, and also to those reported by \citet{dm2013}, \citet{ka2017}, and \citet{df2017}. This suggests that the radial stratification in the initial disks does not affect the gap depth much.
The small differences between Equation \eqref{eq:Surf_depth} (or Equation \eqref{eq:Surf_depth_K15}) and the numerical results at $K\gtrsim 10^3$ for $h_p/r_p\lesssim0.07$ are likely due to the fact that gas responses to such massive planets are highly nonlinear, so that the linear theory of \citet{ka2015a} is not applicable (see also \citealt{ka2015a,ka2017,df2017}).

\subsubsection{Gap Width} \label{sssec:sdw}

\begin{figure}
	\centering
	\includegraphics[width=1.0\linewidth]{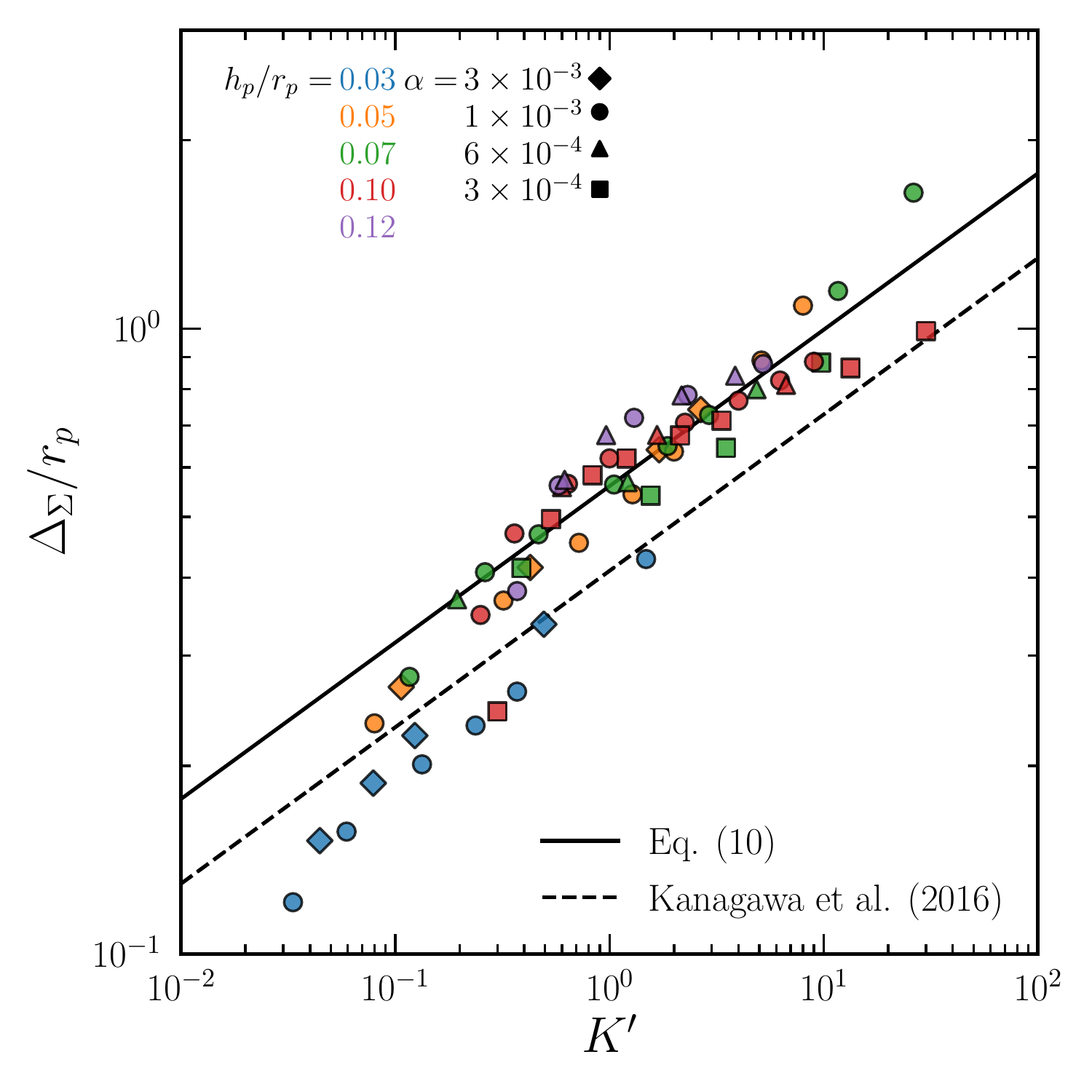}
	\caption{Gap width $\Delta_{\Sigma}$, defined as the radial distance between two points with $\langle\Sigma/\Sigma_0\rangle=1/2$, as a function of  $K'\equiv q^2(h_p/R_p)^{-3}\alpha^{-1}$. The dotted and solid lines draw  Equations \eqref{eq:Surf_width_K15} and \eqref{eq:Surf_width}, respectively.}
	\label{fig:swidth}
\end{figure}

\citet{ka2016} defined the gap width $\Delta_\Sigma$ as the radial distance between two points where $\langle \Sigma/\Sigma_0\rangle=k$, with the threshold value of $k=1/2$, and showed empirically that $\Delta_\Sigma$ depends on a single dimensionless parameter $\Kpr \equiv (h_p/r_p)^2K = q^2 (h_p/r_p)^{-3}\alpha^{-1}$  as
\begin{equation}\label{eq:Surf_width_K15}
\frac{\Delta_\Sigma^{\rm K16}}{r_p} = 0.41 \Kpr^{1/4},
\end{equation}
for uniform disks (see also \citealt{ka2017}).  Figure \ref{fig:swidth} plots $\Delta_\Sigma$ measured in our models with non-uniform disks as a functions of $\Kpr$. To fit the data, we use a functional form same as in Equation \eqref{eq:Surf_width_K15} with power index $1/4$ fixed, and allow a proportional coefficient to vary. The solid line draws our least-square fit
\begin{equation}\label{eq:Surf_width}
\frac{\Delta_\Sigma}{r_p}= 0.56 \Kpr^{1/4},
\end{equation}
which overall gives a wider gap, by about a factor of 1.4, than Equation \eqref{eq:Surf_width_K15} plotted as the dotted line. The discrepancies between $\Delta_\Sigma^{\rm K16}$ and $\Delta_\Sigma$ may arise from the differences in the initial distributions of the disk surface density and temperature.

Figure \ref{fig:swidth} shows that Equation \eqref{eq:Surf_width} overestimates the width at small $K'$. This is expected since the gap width tends to decreases drastically as $\min\langle \Sigma /\Sigma_0\rangle $ approaches the threshold value $1/2$. In fact, the definition of \citet{ka2016} for the gap width cannot be applicable for shallow gaps with $\min\langle\Sigma /\Sigma_0\rangle > 1/2$. Increasing the threshold may alleviate this problem to some extent, but at the expense of increasing the gap width.\footnote{For the threshold density $\langle\Sigma/\Sigma_0\rangle=k$, our numerical results for non-uniform disks are fitted by $\Delta_\Sigma^k/r_p =(0.76k+0.18)\Kpr^{1/4}$, which can be compared with $ \Delta_\Sigma^{{\rm K16},k}/r_p =(0.50k+0.16)\Kpr^{1/4}$ of \citet{ka2017} for uniform disks.} Still, using a fixed threshold density in measuring a gap width is somewhat arbitrary and cannot be applied to all possible gaps.

We thus introduce another definition of a gap width $W_\Sigma$, namely the radial distance between the points where $d\ln\langle\Sigma/\Sigma_0\rangle / d\ln r$ achieves extremum values at both sides of the planet location. The new definition based on the radial gradient of the surface density is motivated to relate the gap width in the surface density to the width in the perturbed velocity profile (see Section \ref{sssec:Deltav}). We try to fit the measured $W_\Sigma$ using various combinations of the input parameters, and find that it is best described by the planet mass normalized by the thermal mass.

\begin{figure}
	\centering
	\includegraphics[width=1.0\linewidth]{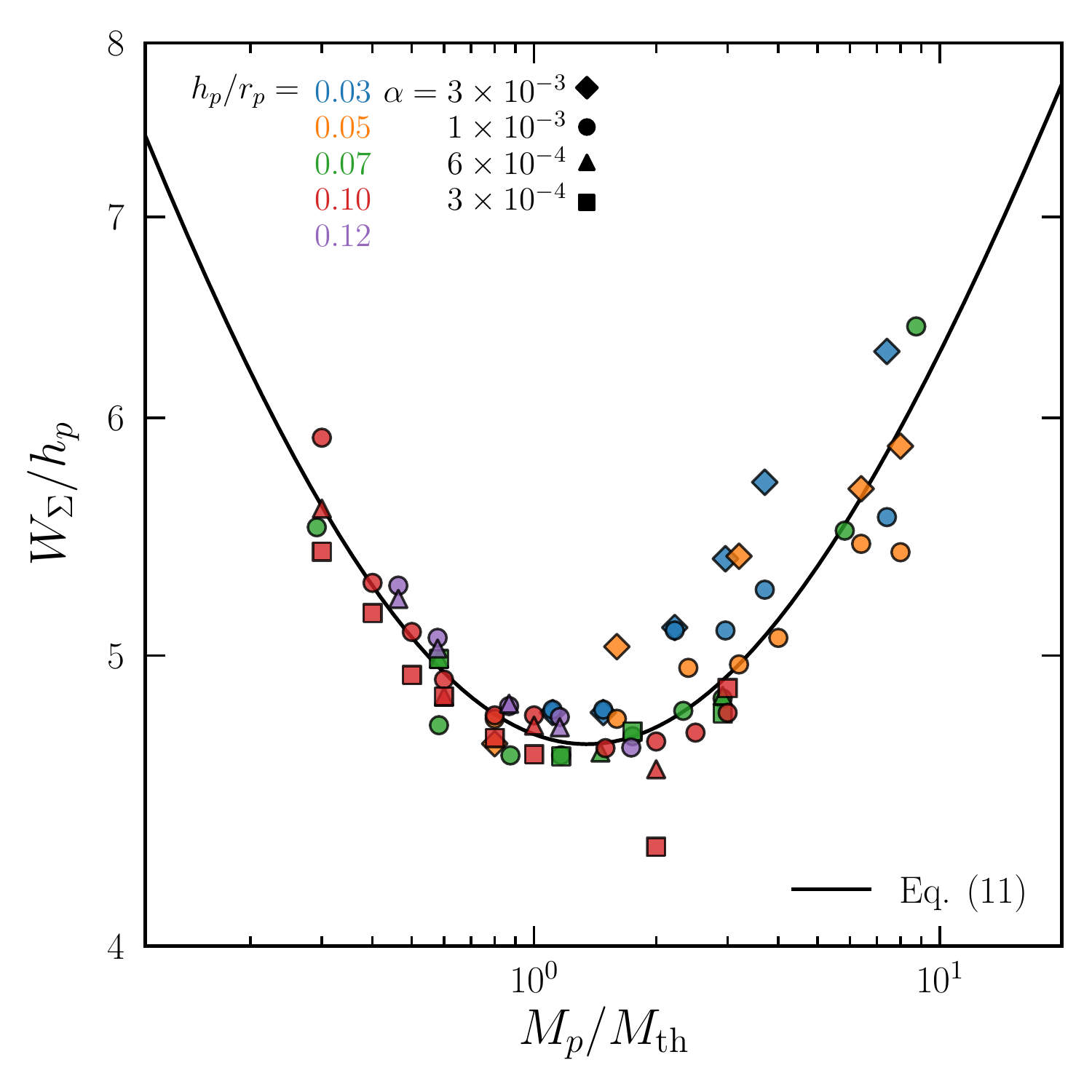}
	\caption{Gap width $W_{\Sigma}$, defined as the radial distance between two extrema in the $d\ln\langle\Sigma/\Sigma_0\rangle/d\ln r$ curve, as a function of $M_p/M\textsubscript{th}$. Note that $W_{\Sigma}/h_p$ increases as $M_p/M_\text{th}$ increases or decreases from about 1.5. The solid line draws our fit, Equation \eqref{eq:Wsurf}.}
	\label{fig:delta2}
\end{figure}

Figure \ref{fig:delta2} plots $W_\Sigma/h_p$ as a function of $M_p/M\textsubscript{th}$. Note that the range of $W_\Sigma/h_p$ is very narrow for the parameters we adopt, with $W_\Sigma\approx 4.7h_p$ on average.
Still, $W_\Sigma/h_p$ depends weakly on the planet mass, such that it increases as $M_p/M_\text{th}$ decreases or increases from about 1.5. This is unlike  $\Delta_{\Sigma}$ which increase monotonically with the planet mass. We fit the data using a linear combination of two power laws in $M_p/M_\text{th}$ with four free parameters (two coefficients and two power indices). Our least-square fit is
\begin{equation}\label{eq:Wsurf}
\frac{W_\Sigma}{h_p}=2.54\left(\frac{M_p}{M_\text{th}}\right)^{-0.43}+2.16\left(\frac{M_p}{M_\text{th}}\right)^{0.39},
\end{equation}
plotted as a solid line on Figure \ref{fig:delta2}. The effect of $\alpha$ on $W_\Sigma$ is almost negligible compared to those of $h_p/r_p$ and $M_p/M_\text{th}$.

The dependence of $W_\Sigma$ on $M_p$ can be understood in terms of the shock formation distance. Since perturbations induced by a low-mass planet are weak even in the regions very close to the planet, they have to travel some distance radially before undergoing nonlinear steepening into shocks. \citet{gr2001} showed that the shock formation distance $l_\text{sh}$ from a planet is given by
\begin{equation}
\frac{l_\text{sh}}{h_p} \approx0.93\left(\frac{\gamma+1}{12/5}\frac{M_p}{M_\text{th}}\right)^{-2/5},
\label{eq:lsh}
\end{equation}
where $\gamma$ is an adiabatic index. Note that the power-law dependence of $l_\text{sh}$ on $M_p/M_\text{th}$ is quite similar to that of $W_\Sigma$ for $M_p/M_\text{th} \lesssim1$ in Equation \eqref{eq:Wsurf}. When $M_p/M_\text{th} \gtrsim1$, on the other hand, perturbations are already nonlinear over a range of radii from the planet location, instantly forming shocks there (e.g., \citealt{do2011}). In this case, the regions (i.e., gap) influenced by shocks become wider for larger $M_p$.

\begin{figure}
	\centering
	\includegraphics[width=1.0\linewidth]{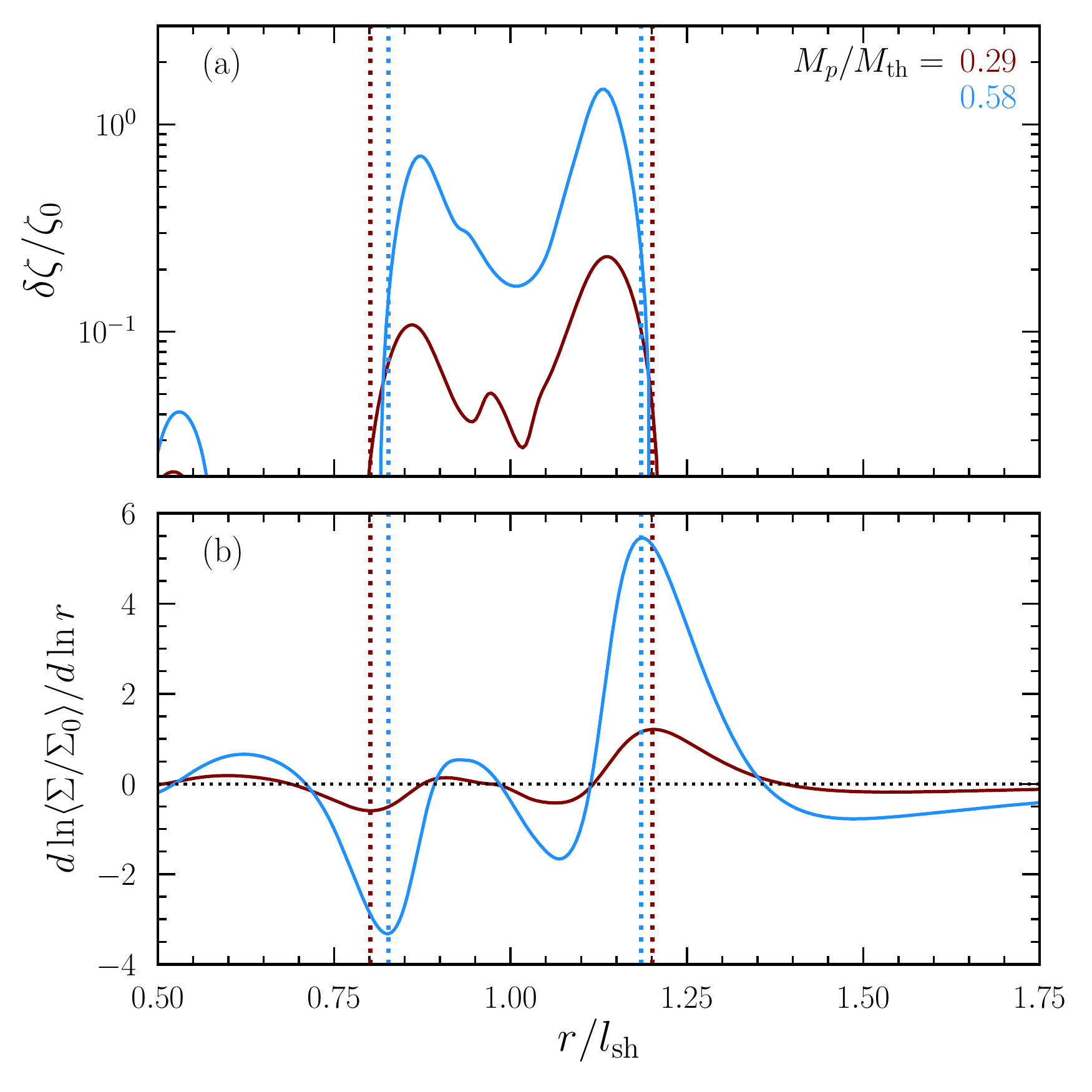}
	\caption{Radial variations of (a) the perturbed potential vorticity $\delta\zeta/\zeta_0$ relative to the initial value $\zeta_0$ and (b) $d\ln\langle\Sigma/\Sigma_0\rangle/d\ln r$ at $t=250t_\text{orb}$ for the models with different mass $M_p/M_\text{th}=0.29$ (brown) and $0.58$ (blue), but with the same $h_p/r_p=0.07$ and $\alpha=3\times10^{-4}$. The vertical dotted lines in (a) and (b) mark the extremum positions of $d\ln\langle\Sigma/\Sigma_0\rangle/d\ln r$, which envelop the regions with significant $\delta\zeta/\zeta_0$.}
	\label{fig:pv}
\end{figure}

To illustrate $W_\Sigma$ is associated with shocks, we calculate the azimuthally-averaged potential vorticity defined as
\begin{equation}
\zeta =\left\langle\frac{|\nabla\times \mathbf{v}|}{\Sigma}\right\rangle.
\end{equation}
Strictly speaking, the potential vorticity in our simulations is not a conserved quantity because the initial disks are not barotropic, so that it is  generated not only by the shock fronts but non-vanishing baroclinic terms.
Since the potential vorticity induced by the baroclinic terms is confined to the corotation regions, its change $\delta\zeta \equiv \zeta-\zeta_0 $ relative to the initial profile $\zeta_0$ away from the corotation is mostly caused by curved shocks. Figure \ref{fig:pv} plots the radial distributions of (a) $\delta\zeta/\zeta_0$ and (b) $d\ln \langle \Sigma/\Sigma_0\rangle/d\ln r$ at $t=250\torb$ for the models with differing $M_p/M_\text{th}=0.29, 0.58$ but with the same $h_p/r_p=0.07$, and $\alpha=3\times10^{-4}$. Apparently, the regions with substantial $\delta\zeta/\zeta_0$ are bounded by the radii
where $d\ln \langle \Sigma/\Sigma_0\rangle/d\ln r$ attains its maximum or minimum, marked by the vertical dotted lines. This is observed for all simulation results which hints that  $W_\Sigma$ for $M_p/M_\text{th}\lesssim1$ can be explained by the shock formation distance.

\begin{figure}
	\centering
	\includegraphics[width=1.0\linewidth]{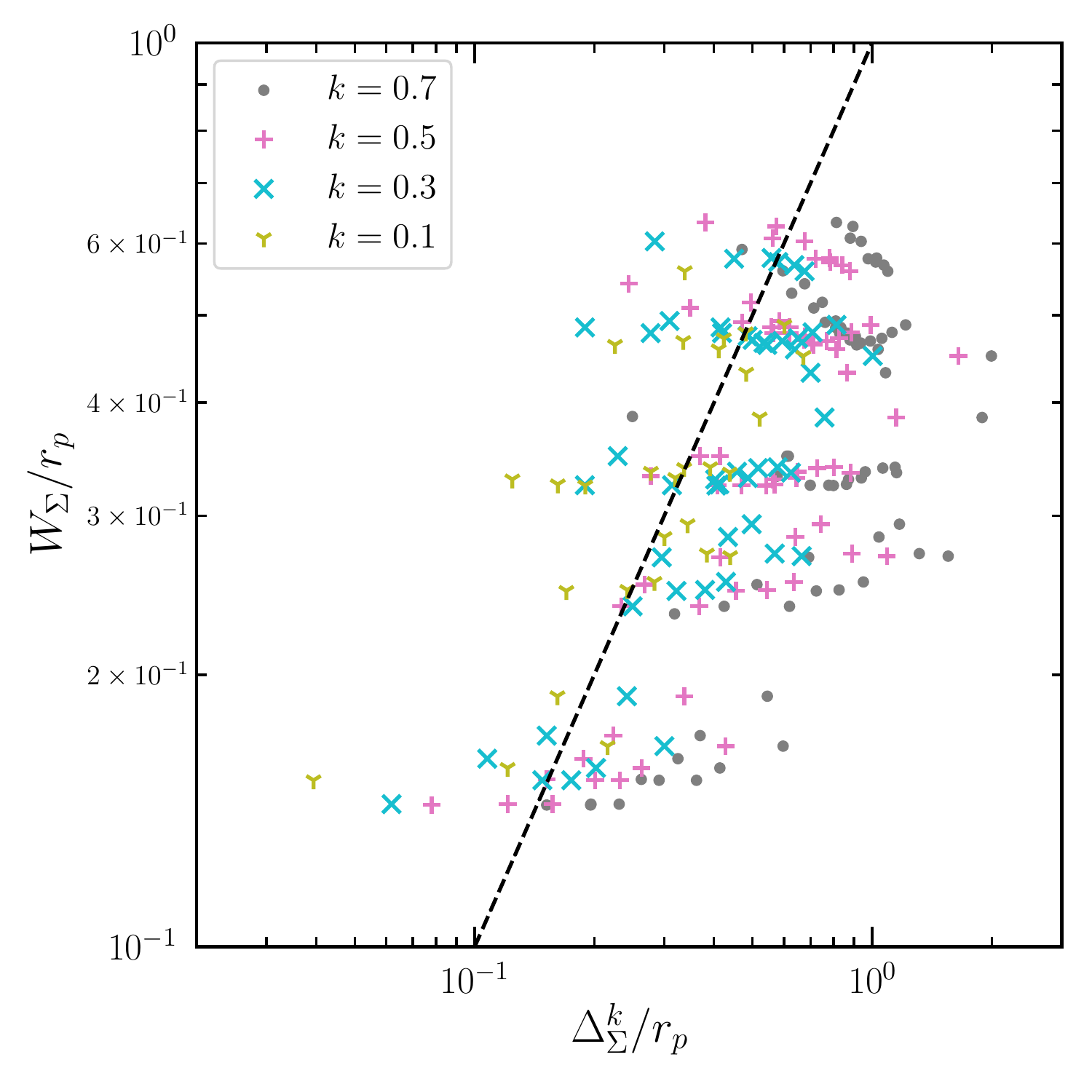}
	\caption{Comparison between $W_\Sigma$ and $\Delta^{k}_{\Sigma}$ with $k=0.1$, $0.3$, $0.5$, and $0.7$. The dashed line corresponds to  $\Delta^{k}_{\Sigma}=W_{\Sigma}$.}
	\label{fig:peaksmin}
\end{figure}

Figure \ref{fig:Surfsnap} hints that the extrema in the radial gradient of $\langle\Sigma/\Sigma_0\rangle$ occur near the bottom of a gap, making $W_\Sigma$ smaller than $\Delta_\Sigma$ with $k=0.5$. Figure \ref{fig:peaksmin} compares $W_{\Sigma}$ and $\Delta^{k}_{\Sigma}$ with differing  threshold $k=0.1$, $0.3$, $0.5$, $0.7$. It is apparent that $\Delta^{k}_{\Sigma}$ is larger for larger $k$. For most cases, $W_{\Sigma}$ is smaller than $\Delta^{k}_{\Sigma}$. Approximately, $W_{\Sigma}$ is similar to $\Delta^{k}_{\Sigma}$ with $k\sim0.3$, indicating that $W_{\Sigma}$ measures the width at the lower part of a gap.

\subsection{Perturbed Rotational Velocity} \label{ssec:rot}

The presence of a planet not only produces a gap in the surface density profile but also induce significant distortion in the rotational velocity $v_\phi$. Figure \ref{fig:Surfsnap}(d) plots the exemplary distribution of the azimuthally-averaged, perturbed velocity $\langle\delta \tilde{v}_\phi\rangle$. Clearly, the radial profile of $\langle \delta \tilde{v}_\phi\rangle$ is nearly anti-symmetric with respect to the planet, with the regions inside (outside) the planet moving slower (faster) than the initial near-Keplerian speed. In this subsection, we quantify the amplitude and width of the perturbed rotational velocity, which rapidly reach a quasi-steady state within $t\sim 10^3\torb$ (Appendix \ref{ap:time}).

\subsubsection{Amplitude of Perturbed Velocity}

\begin{figure}
	\centering
	\includegraphics[width=1.0\linewidth]{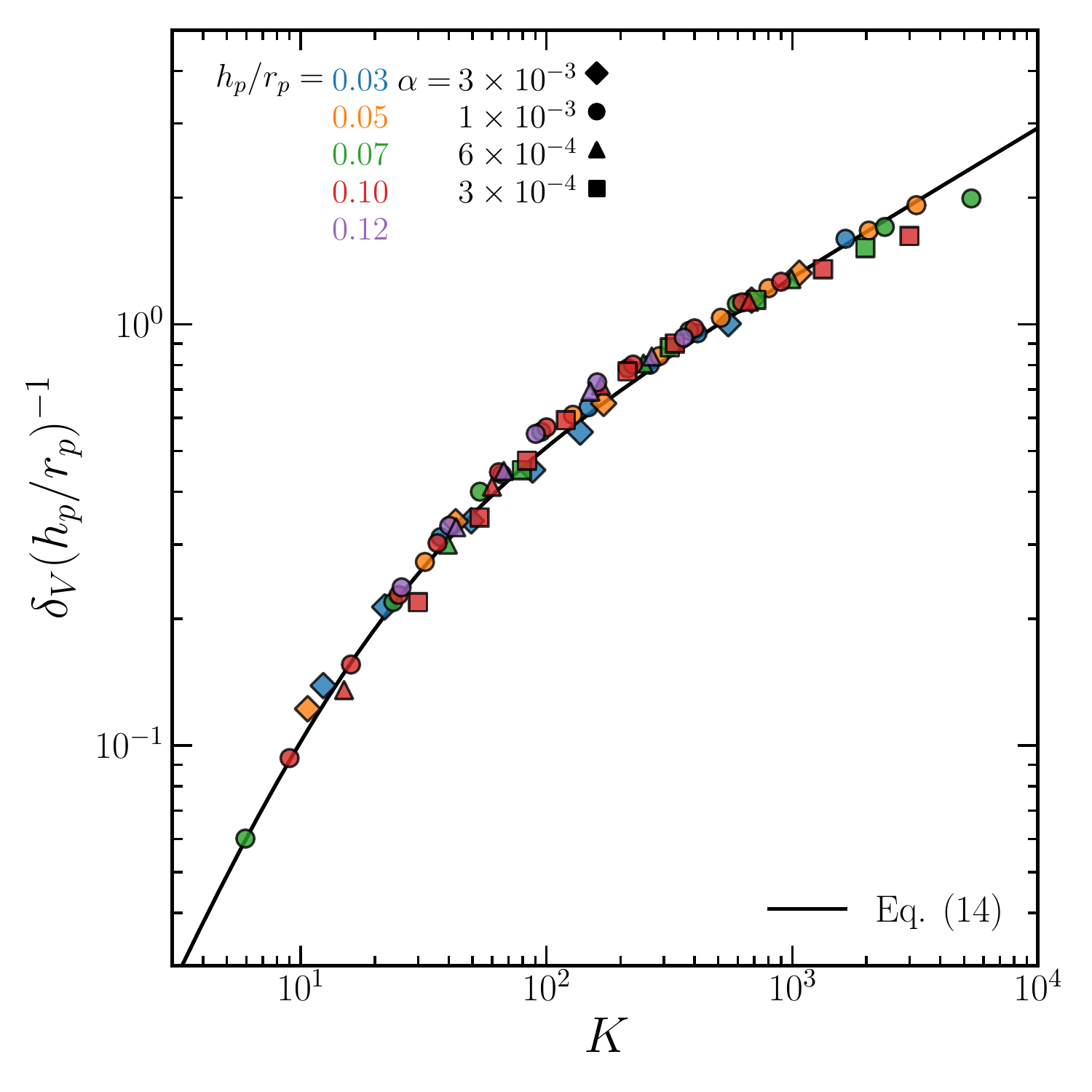}
	\caption{Relationship between $\delta_V(h_p/R_p)^{-1}$ and $K$ for all models. The solid line draws our fit (Equation \eqref{eq:deltav}).}
	\label{fig:vdepth}
\end{figure}

We define the dimensionless amplitude, $\delta_V$, of the perturbed rotational velocity as the difference in $\langle \delta \tilde{v}_\phi\rangle$ between the super-Keplerian peak formed near the outer gap edge and the sub-Keplerian peak near the inner gap edge, as illustrated in Figure \ref{fig:Surfsnap}(d).  We measure $\delta_V$ for all models and plot $\delta_V(h_p/r_p)^{-1}$ as a function of $K$ in Figure \ref{fig:vdepth}. Apparently, the perturbed velocity is larger for a more massive planet in a colder and less diffusive disk. We try to fit $\delta_V(h_p/r_p)^{-1}$ using an inverse of a linear combination of two power laws in $K$ with four free parameters. The solid line draws the resulting least-square fit
\begin{equation}\label{eq:deltav}
\delta_V=\left(\frac{h_p}{r_p}\right)\frac{0.007K^{1.38}}{1+0.06K^{1.03}},
\end{equation}
which is within $~18\%$ of the all measured $\delta_V$. This predicts $\delta_V \propto q^{0.7}(h_p/r_p)^{-0.75}\alpha^{-0.35}$ for $K\gg1$.

The dependence of $\delta_V$ on $h_p/r_p$ and $K$ in Equation \eqref{eq:deltav} results from the fact that the gaps are in hydrostatic equilibrium.
In a quasi-steady state, the force balance in the radial direction reads
\begin{equation}\label{eq:eqSigma}
\frac{v_{\phi}^2}{r}=\frac{GM_*}{r^2}+\frac{1}{\Sigma}\frac{d P}{d r},
\end{equation}
or
\begin{equation}\label{eq:HSE2}
v_{\phi}^2 = v_{\phi,0}^2 + c_s^2 \frac{d\ln(\Sigma/\Sigma_0)}{d\ln r}.
\end{equation}
Since the pressures gradient is negative (positive) near the inner (outer) edge of the gap, the gas there should rotate slower (faster) than the initial velocity in order to maintain an equilibrium (e.g., \citealt{te2018}).
Assuming that the perturbed velocity is much smaller than the initial rotation velocity and that $c_s$ is radially constant, one can show that Equation \eqref{eq:HSE2} reduces to
\begin{equation}\label{eq:dvapp}
\langle \delta\tilde{v}_\phi \rangle \approx \frac{1}{2}\left(\frac{h_p}{r_p}\right)^2\frac{d\ln \langle \Sigma/\Sigma_0\rangle}{d\ln r}.
\end{equation}
We numerically confirm that Equation \eqref{eq:dvapp} holds within 20\% for all models, as evidenced by Figure \ref{fig:Surfsnap}(c) and (d). The deviation becomes larger as the radial range influenced by the planet increases, so that the radial dependence of $c_s$ becomes non-negligible. This proves that the gap width $W_\Sigma$ determined by the extremum positions of $d\ln \langle \Sigma/\Sigma_0\rangle/d\ln r $ traces the sub/super-Keplerian peaks in the velocity profile.

Under the assumption that the $d\ln \langle \Sigma/\Sigma_0\rangle/dr$ profile is anti-symmetric with respect to the planet, Equation \eqref{eq:dvapp} gives
\begin{equation}\label{eq:dvpeak}
\delta_V \sim \left(\frac{h_p}{r_p}\right)^2 \left|\frac{d\ln \langle \Sigma/\Sigma_0\rangle}{d\ln r}\right|_\text{peak}.
\end{equation}
There is no obvious way to calculate $d\ln \langle \Sigma/\Sigma_0\rangle/d\ln r$ at the sub- and super-Keperian peak positions, but it should be related to the gap depth and width, and scale dimensionally as $\propto r_p(1-\delta_{\Sigma})/W_\Sigma$.\footnote{We empirically find $|{d\ln \langle \Sigma/\Sigma_0\rangle}/{d\ln r}|_{\text{peak}}\sim 2.45r_p(1-\delta_\Sigma)^{1.4}/W_\Sigma$ for small $K$.}
Since $\delta_\Sigma$ is a function of $K$ (Figure \ref{fig:smin} and Equation \ref{eq:Surf_depth}) and $W_\Sigma \sim 4.7h_p$ (Figure \ref{fig:delta2}), Equation \eqref{eq:dvpeak} indicates that $\delta_V/(h_p/r_p)$ should be well described by the $K$ parameter alone, consistent with Figure \ref{fig:vdepth}.

\begin{figure}
	\centering
	\includegraphics[width=1.0\linewidth]{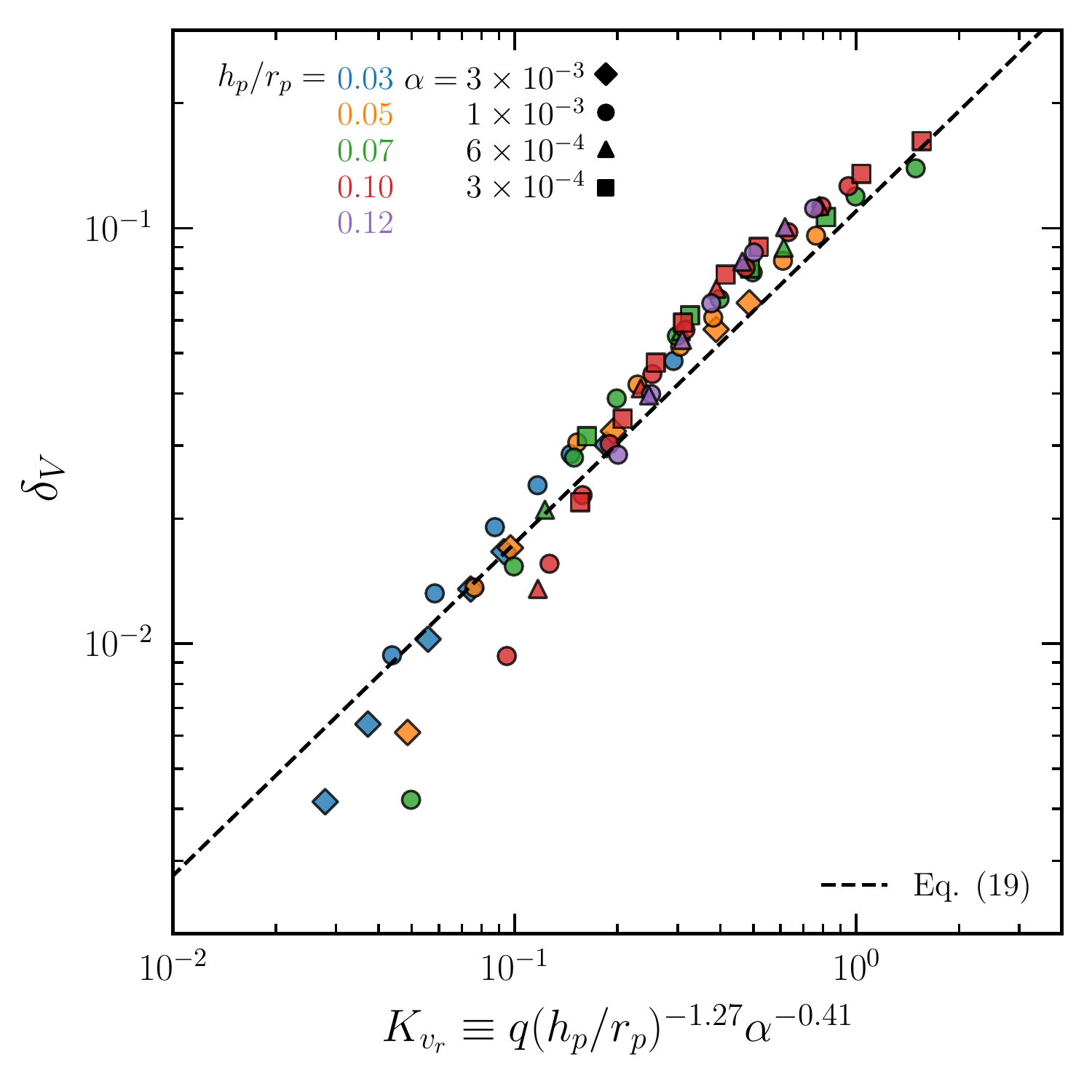}
	\caption{Normalized amplitude $\delta_V$ of the perturbed velocities from our simulations as a function of $K_{v_r}\equiv q(h_p/r_p)^{-1.27}\alpha^{-0.41}$ introduced by \citet{zha2018}. The dotted line draws Equation \eqref{eq:vdkvzha}, the result of \citet{zha2018} in disks without an exponential density cutoff.}
	\label{fig:vdcomp}
\end{figure}

\citet{zha2018} also studied the dependence on the disk parameters of the amplitude of perturbed velocities by using disk models similar to ours but without the exponential cutoff in the initial density distribution (Equation \ref{eq:sigmao}).
They found that the amplitude of the sub/super-Keplerian peaks in the velocity profile is well fitted by
\begin{equation} \label{eq:vdkvzha}
\delta_V=0.11 K_{v_r}^{0.80},
\end{equation}
where  $K_{v_r}\equiv q(h_p/r_p)^{-1.27}\alpha^{-0.41}$. Figure \ref{fig:vdcomp} plots $\delta_V$ measured from our simulations as a function of $K_{v_r}$. Our measured $\delta_V$ agrees, mostly within 30\%, with Equation \eqref{eq:vdkvzha} plotted as a dotted line, with a small discrepancy between the two caused most likely by the difference in the initial density distribution.

\subsubsection{Width of Perturbed Regions} \label{sssec:Deltav}

\begin{figure}
	\centering
	\includegraphics[width=1.0\linewidth]{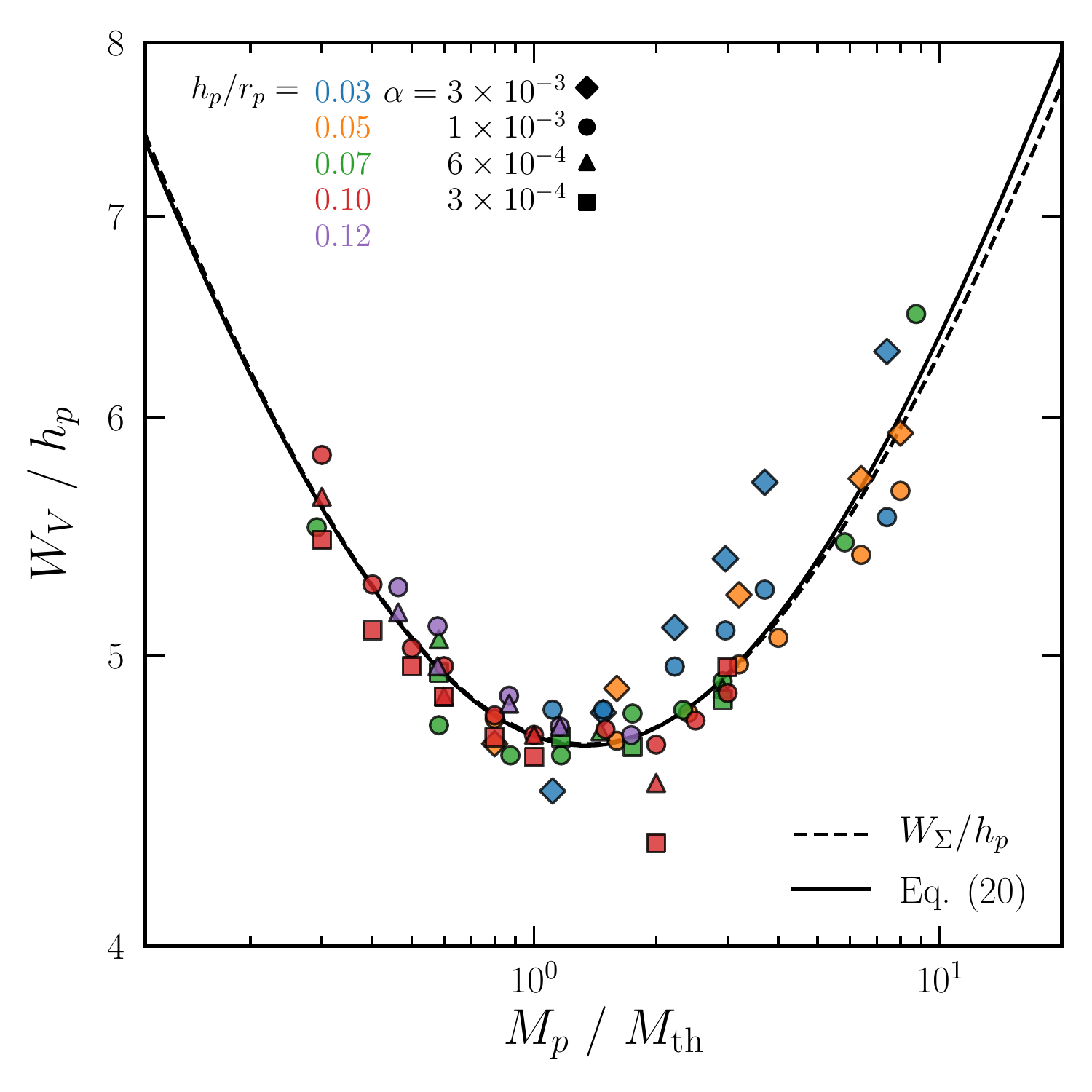}
	\caption{Dependence on $M_p/M_\text{th}$ of the normalized width $W_V/h_p$ of the regions with significant perturbed velocities. Note that $W_V$ increases as $M_p/M_\text{th}$ decreases or increases from $\sim 1.5$. The solid line is our fit (Equation \ref{eq:Wvel}), while the dashed line draws Equation \eqref{eq:Wsurf} for $W_\Sigma/h_p$. }
	\label{fig:vwidth}
\end{figure}

We define the width, $W_V$, of the regions with significant perturbed velocities as the radial distance between the super- and sub-Keplerian peaks in the $\langle \tilde{v}_\phi \rangle$ profile, as indicated in Figure \ref{fig:Surfsnap}(d). As mentioned above, $W_V$ would be similar to $W_\Sigma$ if the disks have constant temperature, but the non-uniform temperature distribution makes them slightly different from each other. Figure \ref{fig:vwidth} plots $W_V/h_p$ as a function of $M_p/M_{\text{th}}$.  Similarly to $W_\Sigma/h_p$, we fit the numerical results for $W_V/h_p$ using a linear combination of two power laws in $M_p/M_\text{th}$ with four free parameters. Our least-square fit is
\begin{equation}\label{eq:Wvel}
\frac{W_V}{h_p}=2.66\left(\frac{M_p}{M_{\text{th}}}\right)^{-0.41}
+2.04\left(\frac{M_p}{M_{\text{th}}}\right)^{0.42},
\end{equation}
plotted as a solid line. For comparison, we overplot Equation \eqref{eq:Wsurf} for $W_\Sigma/h_p$ as a dashed line, which is very close to Equation \eqref{eq:Wvel}, suggesting that Equation \ref{eq:dvapp} is a good approximation. As in $W_\Sigma$, the range of $W_V$ is very narrow for the parameters adopted, with $W_V\approx 4.7h_p$ on average. This is in agreement with \citet{zha2018} who found $W_V\approx 4.4h_p$.
Again, $W_V$ follows a power law for $M_p/M_\text{th}< 1$, with an index very close to $-0.4$, suggesting that the width of the perturbed regions is determined by the shock formation distance for a low-mass planet.

\section{Discussion} \label{sec:Discussion}


So far, we have provided the quantitative dependence on the input parameters of the gap depth $\delta_{\Sigma}$ and width $\Delta_{\Sigma}$ or $W_\Sigma$ in the perturbed density profile as well as the amplitude $\delta_V$ and spatial range $W_V$ of the perturbed velocities. Most observations with ALMA trace dust rather than gas in the disks, while the gap parameters measured in the present work are for the surface density and velocity distributions in the gaseous component. One thus needs to convert dust-continuum emissions to surface density maps to obtain $\delta_{\Sigma}$ and $\Delta_\Sigma$ \citep{df2017}, but the conversion process can easily be affected by the dust-to-gas ratio, dust properties, chemical effects, etc., which are quite uncertain \citep{be2013,mi2017}.  However, $\delta_V$ and $W_V$ are relatively free of the conversion problem because one can directly measure the perturbed rotational velocities in the gaseous disks \citep{pi2018,te2018}.

Still, the relations presented in the preceding section are based on the 2D simulations, while observed rotational velocities are derived at the emission surface of a certain tracer, which is typically above the disk midplane. To estimate the effects of the vertical disk stratification, we follow \citet{da2003} and \citet{an2012} to consider a thermally-stratified, axisymmetric disk in the $r$--$z$ plane with temperature distribution
\begin{align}
 T(r,z)=
	\begin{cases}
	\footnotesize T_a, &\text{for $z\geq z_q$,} \\
	\footnotesize T_a +(T_0 - T_a)\displaystyle\cos^4\left(\frac{\pi z}{2z_q}\right), &\text{for $ z<z_q$,}
	\end{cases}
\end{align}
where $T_0(r)$ is the midplane temperature (Equation \ref{eq:eqT}) and $T_a(r)$ is the temperature of the disk atmosphere at $z\geq z_q\equiv3h(r)$. We consider three models with $T_a=nT_0$ for $n=1$, 2, and 3: $n=1$ corresponds to an isothermal disk in $z$.

The condition of hydrostatic equilibrium along the $z$-direction requires that the mass density $\rho$ obeys
\begin{equation}
\frac{\rho(r,z)}{\rho(r,0)}=\frac{c_s^2(r,0)}{c_s^2(r,z)}\exp\left[-\int_0^{z} \frac{1}{c_s^2}\frac{GM_*z'}{(r^2+z'^2)^{3/2}}dz'  \right].
\end{equation}
The force balance along the radial direction (cf.~Equation \ref{eq:eqSigma})
allows us to calculate the equilibrium rotational velocity $v_\phi(r,z)$ in the $r$--$z$ plane.

\begin{figure*}
	\centering
	\includegraphics[width=1.0\linewidth]{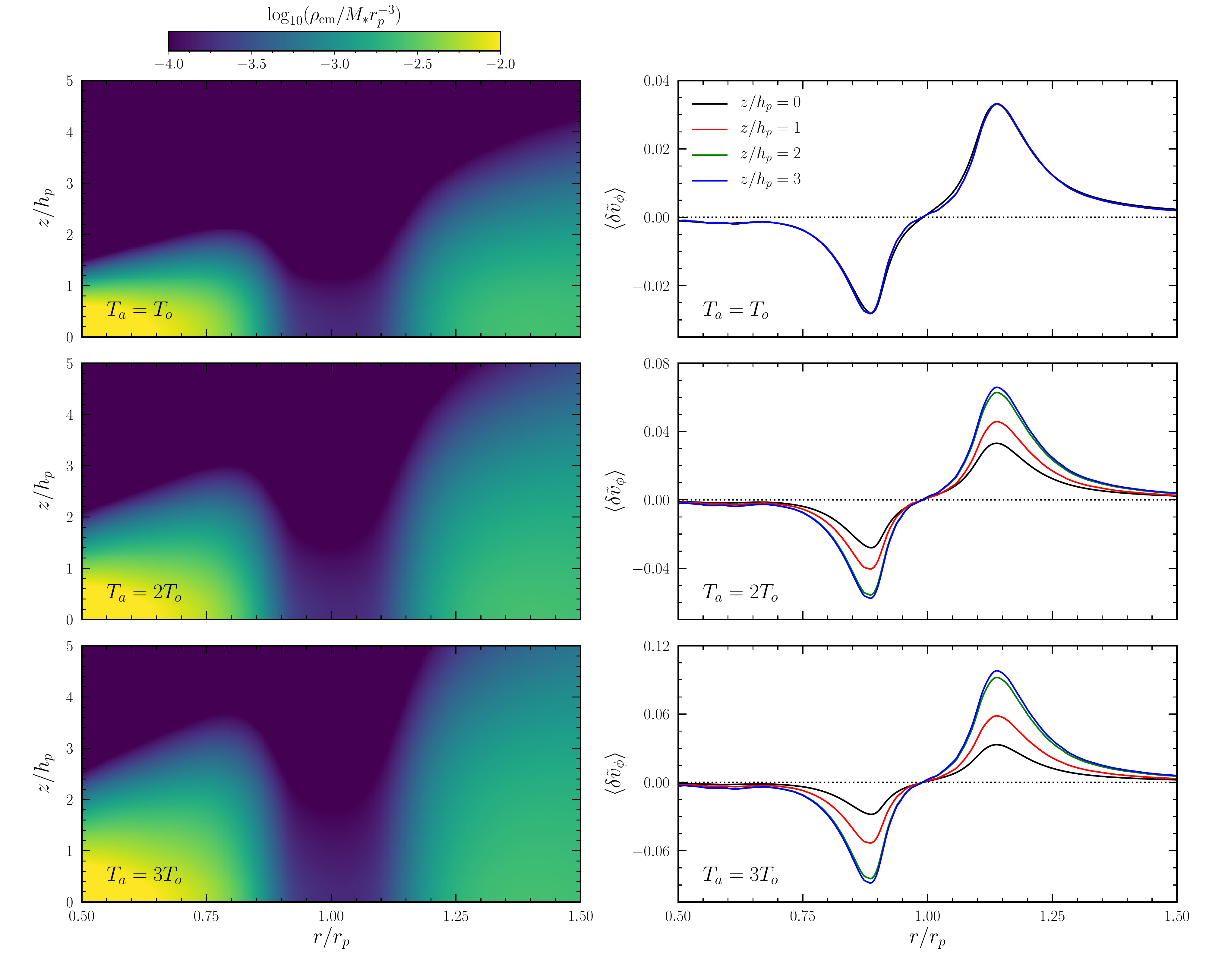}
	\vspace*{-5mm}
	\caption{(left) Distributions in the $r$--$z$ plane of the gas density in vertically stratified disks with $T_a/T_0=1$, 2, 3 from top to bottom, for the model with $h_p/r_p=0.05$, $M_p/M_*=5\times10^{-4}$, and $\alpha=1.0\times10^{-3}$. (right)  Perturbed rotational velocities normalized to the initial velocities $\langle \delta \tilde{v}_\phi\rangle$ at $z/h_p=0$ (black), 1 (red), 2 (green), and 3 (blue). While the width $W_V$ is almost unchanged with $z$, the amplitude $\delta_V$ becomes larger for larger $z$ and $T_a/T_0$ to maintain a hydrostatic equilibrium in the radial direction.}
	\label{fig:res_rz}
\end{figure*}

The left panels of Figure \ref{fig:res_rz} plot the logarithm of the mass density in the $r$--$z$ plane of the stratified disk models for $T_a/T_0=1$, 2, 3 from top to bottom. The case with $h_p/r_p=0.05$, $M_p/M_*=5\times10^{-4}$, and $\alpha=1.0\times10^{-3}$ is chosen. The right panels plot the radial distribution of $\langle \delta \tilde{v}_\phi\rangle$ at certain heights $z=h_p$, $2h_p$, $3h_p$ in comparison with the 2D results (i.e., at $z=0$). Note that the width $W_V$ of the perturbed regions is almost independent of $z$. However, the amplitude $\delta_V$ of the perturbed velocity is boosted significantly as $z$ increases in a thermally-stratified disk, and {the amount of the boost is proportional to $T_a/T_0$ through the pressure gradient in the radial direction. This suggests that the 2D results may not be applicable to optically-thick disks for which emission comes from high-$z$ regions. In this case, it is desirable to run three-dimensional simulations with radiation transfer included in order to incorporate the vertical temperature distribution as well as gas mixing induced by a planet.

With the caveat that our empirical relations are based on simulations with a single planet, we apply our empirical results to the observed rotational velocity of the C$^{18}$O(2-1) emission from the HD 163296 disk presented by \citet{te2018}. Because the inner regions of the disk suffer from insufficient spatial resolution in precisely measuring the rotational velocity \citep{te2018}, we focus on the outer two gaps.  By using the disk temperature model of \citet{fl2017}, $\alpha = 10^{-3}$ and assuming that C$^{18}$O traces $z/r \sim 0.15$ \citep{te2018}, the observed gap depth corresponds to $\delta_V \sim 0.020$ and $\sim 0.034$ for the middle and outermost gaps at $r=100\,$AU and $165\,$AU, respectively.\footnote{These gap locations are based on the pre-Gaia distance of $122\rm\,pc$ to HD 163296 \citep{van1997}. The corresponding Gaia distance is $101.5\rm\,pc$ \citep{gaia2018}.} Our empirical relation, Equation \eqref{eq:deltav}, then gives $M_p\sim 0.38$--$0.52\,M_J$ and $\sim0.76$--$1.11\,M_J$ allowing for 18\% uncertainties, respectively, for middle and outermost gaps, where $M_J$ is the Jupiter mass. We can also place constraints on the planet mass using the observed gap width in rotation velocities. For the middle gap at 100 AU, we obtain $W_V/h_p = 4.4$ for the observed width $W_V \sim 25$ AU and $h_p/r_p \sim 0.057$. This suggests that the planet mass could be close to the thermal mass ($M_{\text{th}} = 0.42 M_J$), consistent with the above estimate using the gap depth relation, although we should point out that $W_V/h_p = 4.4$ is smaller than the minimum of our best fit (4.7; Equation \ref{eq:Wvel}). We conjecture that this is presumably because the interaction between multiple planets could have modified the gap shape; for the middle gap in particular, it is quite possible that the gap could become narrower than otherwise, as the disk gas is pushed by both inner and outer planets. For the outer gap at 165 AU, the outer $\delta_v$ peak location cannot be well defined in the observations because the signal-to-noise ratio of the CO data drops in the outer disk (see Figure 5 of \citealp{te2018}). Depending on the exact peak location, the gap width ranges from 55 AU to 75 AU, which correspond to $W_V/h_p = 4.8$--$6.6$. Because this covers a broad range of planet mass (Figure \ref{fig:vwidth}), we cannot make a meaningful estimate based on the data presented in \citet{te2018}.

It is worth noting that the above inferred planet masses have to be regarded as a lower limit to the actual planet mass, because the observed rotation velocity deviations are smoothed with a synthesized ALMA beam and thus have to be smaller than the intrinsic values. In \citet{te2018}, they obtained simulated CO rotation velocities by convolving the raw velocity field from a two-dimensional planet-disk interaction simulation with a synthesized ALMA beam ($0.26'' \times 0.18''$). Taking into the beam convolution account, they needed 1.0 and 1.3 $M_J$ planets to reproduce the observation, a factor of 2.3 and 1.5 larger than our estimates, respectively. These discrepancies between the planet mass with and without beam convolution suggest that one should be careful when applying our relation for $\delta_V$ directly to observations.

As a measure of gap width in real observations, \citet{zha2018} suggested the width $\Delta_\Sigma^\text{Z18}$ normalized by the location of the outer gap edge instead of the planet position $r_p$ since the latter is hardly constrained observationally. Assuming that the density gap is symmetric with respect to the planet, one can express $\Delta_\Sigma^\text{Z18}$ in terms of $\Delta_\Sigma$ as
\begin{equation}\label{eq:wapprox}
\Delta_\Sigma^\text{Z18}\approx\frac{\Delta_\Sigma/r_p}{1+\Delta_\Sigma/(2r_p)},
\end{equation}
and a similar expression for the spatial width of the regions with significant velocity perturbations. Using our simulations, we check that Equation \eqref{eq:wapprox} is accurate within 8\%, suggesting that the $\langle \Sigma/\Sigma_0\rangle$ is nearly symmetric relative to the planet.


In this paper, we have explored various gap properties produced by planets using simple numerical simulations. There are certainly many caveats that need to be improved in future studies. Our models consider only gaseous disks and neglect the effects of dust. A dust-gas mixture is prone to
streaming instability \citep{yg2005} and the gap structure can be altered by the frictional feedback of dust when a sufficient amount of dust is trapped at the edge of the gap \citep{ka2018}.  In addition, our simulations do not allow for planet migration by taking a fixed circular orbit. \citet{na2019} showed that the number and shape of gaps depend on the migration speed of a planet and the drift speed of dust. Also, increasing an inclination angle of the planet orbit relative to the disk midplane tends to make a gap shallower \citep{zh2018}.

Our models adopt viscous disks with $\alpha\geq 3\times 10^{-4}$, so that we are unable to explore multiple gaps launched by a single planet commonly found in low viscosity disks with $\alpha<10^{-4}$ \citep{do2017,ba2017}. A number of studies investigated the spacing, depth, and number of gaps \citep{do2018b,zha2018}, but other gap parameters such as the gap depth in the surface profile and the properties of the associated velocity have yet to be explored for multiple gaps. Comparison of the properties between primary and secondary gaps would help distinguish whether observed multiple gaps are launched by a single or multiple planets.

Finally, our simulations do not include the effects of magnetic fields that may be pervasive in protoplanetary disks. Previous work that ran magnetohydrodynamic simulations of protoplanetary disks reported that gap structure can be changed considerably by magnetic fields \citep{wi2003,np2003,ur2011,zh2013}. The presence of magnetic fields tends to make gaps wider compared to unmagnetized counterparts, there is no consensus on the effect of magnetic fields on the gap depth. For instance, \citet{wi2003} with toroidal fields reported that turbulence driven by magnetorotational instability makes the gaps shallower, while \citet{np2003} and \citet{zh2013} with initial poloidal fields found turbulence makes the gaps deeper that the hydrodynamic cases. It is uncertain whether the discrepancies in the results with magnetic fields are due to filed geometry, disk structure, numerical methods, or resolution. This issue will be addressed by comparing the results of simulations in which one one parameter is varied, while the other parameters are fixed.

\section{Summary} \label{sec:Summary}

We run 2D hydrodynamic simulations of protoplanetary disks with an embedded planet to study the properties of gaps in the surface density profile and the perturbed rotational velocity induced by the planet. We assume that the disks are razor thin, locally isothermal, unmagnetized, non-self-gravitating, and  non-uniform in the radial direction. To investigate various situations, we vary the mass ratio $q=M_p/M_*$ of a planet to a central star, the ratio $h_p/r_p$ of the disk scale height to the orbital radius of the planet, and the viscosity parameter $\alpha$. We measure the gap depth and width in the surface density and velocity profiles after $t=10^4\torb$ when a system reaches a quasi-steady state, and fit them using various combinations of the input parameters.
Our main results can be summarized as follows.

\begin{enumerate}
\item The gap depth $\delta_{\Sigma}$ in the surface density profile in our non-uniform disks is well described by the $K=q^2(h_p/r_p)^{-5}\alpha^{-1}$ parameter introduced by \citet{ka2015a} as $\delta_{\Sigma}=(1+0.046K)^{-1}$ (see Equation \ref{eq:Surf_depth} and Figure \ref{fig:smin}), which is very close to the well-known relation $\delta_{\Sigma}\approx(1+0.04K)^{-1}$ of \citet{ka2015a} for uniform disks.

\item The gap width $\Delta_{\Sigma}$ defined as the radial distance between two points with $\langle \Sigma/\Sigma_0\rangle=1/2$ behaves as $\Delta_\Sigma/r_p=0.56K^\prime$ (see Equation \ref{eq:Surf_width} and Figure \ref{fig:swidth}), where $K^\prime=q^2(h_p/r_p)^{-3}\alpha^{-1}$ is a dimensionless parameter introduced by \citet{ka2016}. Gaps in our non-uniform disks are wider than those in uniform disks by a factor of $\sim1.4$.

\item An alternative gap width $W_\Sigma$ based on the radial gradient of the surface density profile has a minimum $W_\Sigma\approx 4.7h_p$ at $M_p\sim M_\text{th}$, while depending weakly on $M_p/M_\text{th}$ as $W_\Sigma/h_p=2.54(M_p/M_\text{th})^{-0.43}+2.16(M_p/M_\text{th})^{0.39}$, with $M_\text{th}$ being the thermal mass (see Equation \ref{eq:Wsurf} and Figure \ref{fig:delta2}). The power-law dependence of $W_\Sigma$ on $M_p/M_\text{th}<1$ suggests that the gap formation involves nonlinear steepening of perturbations into shocks for low-mass planets.

\item The dimensionless amplitude of the perturbed rotational velocity $\delta_V$, defined as the difference between the positive peak and the negative peak in the $\langle (v_\phi-v_{\phi,0})/v_{\phi,0}\rangle$ profile can be parameterized by $K$ as $\delta_V(h_p/r_p)^{-1}=0.007K^{1.38}/(1+0.06K^{1.03})$ (see Equation \ref{eq:deltav} and Figure \ref{fig:vdepth}). The perturbed rotational velocity is directly related to the radial gradient of the surface density profile via Equation \eqref{eq:dvapp}.

\item Similarly to $W_\Sigma$, the spatial width $W_V$ of the regions with significant velocity perturbations is minimized to $W_V\approx 4.7h_p$ at $M_p/M_\text{th}\sim1$, and depends weakly on $M_p/M_\text{th}$ as $W_V/h_p=2.66(M_p/M_\text{th})^{-0.41}+2.04(M_p/M_\text{th})^{0.42}$ (see Equation \ref{eq:Wvel} and Figure \ref{fig:vwidth}). This suggests that the width of the perturbed regions is determined by the shock formation distance for a low-mass planet with $M_p/M_\text{th}<1$.

\end{enumerate}

These parameterized gap properties can be applied to observations to infer the planet mass and orbital radius as well as the disk properties that are difficult to constrain observationally, for gaps produced by planets.

\acknowledgements

We are grateful to an anonymous referee for an insightful report. This work was supported by grant 2017R1A4A1015178 of the National Research Foundation of Korea. The computation of this work was supported by the Supercomputing Center/Korea Institute of Science and Technology Information with supercomputing resources including technical support (KSC-2018-CHA-0047).

\appendix

\section{Effects of the Indirect Term}\label{ap:indirect}

To illustrate the effects of the indirect term ($\Phi_{i}\equiv GM_pr\cos(\phi-\phi_p)/r_p^2$) arising from the motion of the central star relative to the center of mass, we run a simulation for $h_p/r_p=0.07$, $\alpha=1\times10^{-3}$ and $M_p=0.4\,M_J$ by including the indirect term in the momentum equations. Figure \ref{fig:indirect} compares the radial distributions of the normalized surface density $\langle\Sigma/\Sigma_0\rangle$ and the normalized perturbed velocity $\langle\delta \tilde{v}_\phi\rangle$ averaged over the azimuthal direction and time $t=10^4$--$10100\torb$ between the cases with and without the indirect term.
The indirect term make almost negligible (less than 1\%) changes to the gap profiles, consistent with the results of \citet{ka2017} that the indirect term do not significantly contribute to the angular momentum flux of waves generated by a planet. This confirms that one can ignore the indirect term in measuring the gap properties.

\begin{figure}
	\centering \includegraphics[width=0.9\linewidth]{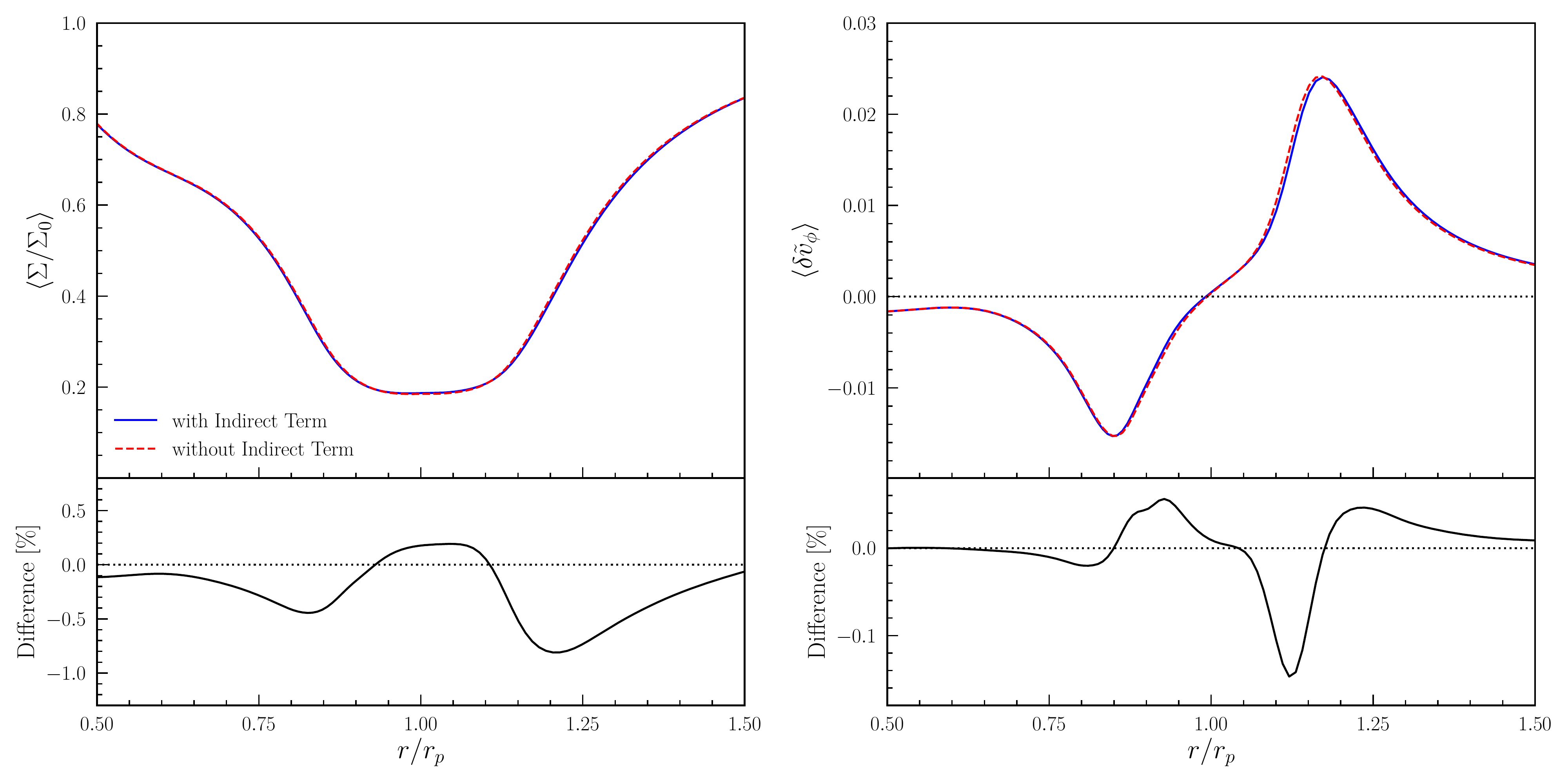}
	\caption{Radial distributions of (left) $\langle\Sigma/\Sigma_{0}\rangle$ and (right) $\langle \delta \tilde{v}_\phi \rangle$ for $h_p/r_p=0.07$, $\alpha=1\times10^{-3}$ and $M_p=0.4M_J$. The blue and red lines correspond to the cases with and without the indirect term, respectively. The lower panels draw the differences between the two cases.}
	\label{fig:indirect}
\end{figure}

\section{Model Parameters and Measured Gap Properties}\label{ap:tbl}

Table \ref{tb:param} lists the model parameters and gap properties measured from the radial distributions of $\langle\Sigma/\Sigma_0\rangle$ and
$\langle \delta \tilde{v}_\phi\rangle=\langle (v_\phi-v_{\phi,0})/v_{\phi,0} \rangle$ averaged over the azimuthal direction and over $t=10000\torb$--$10100\torb$. Columns (1)--(3) give the disk aspect ratio $h_p/r_p$, viscosity parameter $\alpha$, and mass ratio $q=M_p/M_*$, respectively. Columns (4)--(6) give the depth $\delta_\Sigma$ from the bottom in the $\langle\Sigma/\Sigma_0\rangle$ distribution, width $\Delta_\Sigma$ defined by the radial distance between two points where $\langle\Sigma/\Sigma_0\rangle=0.5$, and width $W_\Sigma$ defined by the distance between minimum and maximum points in the $d\langle\Sigma/\Sigma_0\rangle/dr$ distribution. Columns (7) and (8) give the depth $\delta_V$ and width width $W_V$ defined by the difference in $\langle \delta \tilde{v}_\phi\rangle$ and the radial distance between the super- and sub-Keplerian peaks. All quantities are dimensionless.

\startlongtable
\begin{deluxetable}{c|c|c|ccccc}
\tablehead{
\colhead{$h_p/r_p$} & \colhead{$\alpha$} & \colhead{$q$} & \colhead{$\delta_{\Sigma}$} & \colhead{$\Delta_{\Sigma}/r_p${\tablenotemark{a}} } & \colhead{$W_{\Sigma}/r_p$} & \colhead{$\delta_V$} & \colhead{$W_V/r_p$} \\
\colhead{(1)} & \colhead{(2)} & \colhead{(3)} & \colhead{(4)} & \colhead{(5)} &
\colhead{(6)} & \colhead{(7)} & \colhead{(8)}
}
\tablecaption{Model Parameters and Simulation Outcomes\label{tb:param}}
\startdata
\multirow{10}{*}{0.03}	& \multirow{6}{*}{1$\times10^{-3}$}	& 3.0$\times10^{-5}$	& 3.99$\times10^{-1}$	& 0.121			& 0.144	& 9.37$\times10^{-3}$	& 0.144 \\
						& 									& 4.0$\times10^{-5}$	& 2.87$\times10^{-1}$	& 0.157			& 0.144	& 1.32$\times10^{-2}$	& 0.144 \\
						& 									& 6.0$\times10^{-5}$	& 1.63$\times10^{-1}$	& 0.201			& 0.153	& 1.91$\times10^{-2}$	& 0.149 \\
						& 									& 8.0$\times10^{-5}$	& 9.72$\times10^{-2}$	& 0.232			& 0.153	& 2.41$\times10^{-2}$	& 0.153 \\
						& 									& 1.0$\times10^{-4}$	& 5.91$\times10^{-2}$	& 0.263			& 0.158	& 2.86$\times10^{-2}$	& 0.158 \\
						& 									& 2.0$\times10^{-4}$	& 5.75$\times10^{-3}$	& 0.428			& 0.167	& 4.79$\times10^{-2}$	& 0.167 \\
\cline{2-8}
						& \multirow{4}{*}{3$\times10^{-3}$}	& 3.0$\times10^{-5}$	& 5.77$\times10^{-1}$	& \textendash\	& 0.144	& 4.16$\times10^{-3}$	& 0.135 \\
						& 									& 4.0$\times10^{-5}$	& 4.76$\times10^{-1}$	& 0.078			& 0.144	& 6.40$\times10^{-3}$	& 0.144 \\
						& 									& 6.0$\times10^{-5}$	& 3.35$\times10^{-1}$	& 0.152			& 0.153	& 1.02$\times10^{-2}$	& 0.153 \\
						& 									& 8.0$\times10^{-5}$	& 2.42$\times10^{-1}$	& 0.188			& 0.162	& 1.35$\times10^{-2}$	& 0.162 \\
						& 									& 1.0$\times10^{-4}$	& 1.76$\times10^{-1}$	& 0.224			& 0.171	& 1.66$\times10^{-2}$	& 0.171 \\
						& 									& 2.0$\times10^{-4}$	& 3.84$\times10^{-2}$	& 0.337			& 0.189	& 3.01$\times10^{-2}$	& 0.189 \\
\hline
\multirow{12}{*}{0.05}	& \multirow{7}{*}{1$\times10^{-3}$}	& 1.0$\times10^{-4}$	& 3.84$\times10^{-1}$	& 0.234			& 0.238	& 1.36$\times10^{-2}$	& 0.238 \\
						& 									& 2.0$\times10^{-4}$	& 1.51$\times10^{-1}$	& 0.368			& 0.238	& 3.05$\times10^{-2}$	& 0.234 \\
						& 									& 3.0$\times10^{-4}$	& 7.24$\times10^{-2}$	& 0.455			& 0.248	& 4.20$\times10^{-2}$	& 0.239 \\
						& 									& 4.0$\times10^{-4}$	& 3.71$\times10^{-2}$	& 0.543			& 0.248	& 5.18$\times10^{-2}$	& 0.248 \\
						& 									& 5.0$\times10^{-4}$	& 1.96$\times10^{-2}$	& 0.636			& 0.253	& 6.09$\times10^{-2}$	& 0.253 \\
						& 									& 8.0$\times10^{-4}$	& 3.06$\times10^{-3}$	& 0.889			& 0.272	& 8.35$\times10^{-2}$	& 0.270 \\
						& 									& 1.0$\times10^{-3}$	& 6.58$\times10^{-4}$	& 1.088			& 0.271	& 9.59$\times10^{-2}$	& 0.284 \\
\cline{2-8}
						& \multirow{5}{*}{3$\times10^{-3}$}	& 1.0$\times10^{-4}$	& 5.63$\times10^{-1}$	& \textendash\	& 0.234	& 6.11$\times10^{-3}$	& 0.234 \\
						& 									& 2.0$\times10^{-4}$	& 3.13$\times10^{-1}$	& 0.267			& 0.252	& 1.70$\times10^{-2}$	& 0.244 \\
						& 									& 4.0$\times10^{-4}$	& 1.24$\times10^{-1}$	& 0.415			& 0.270	& 3.25$\times10^{-2}$	& 0.262 \\
						& 									& 8.0$\times10^{-4}$	& 2.33$\times10^{-2}$	& 0.640			& 0.284	& 5.71$\times10^{-2}$	& 0.286 \\
						& 									& 1.0$\times10^{-3}$	& 1.06$\times10^{-2}$	& 0.742			& 0.294	& 6.61$\times10^{-2}$	& 0.297 \\
\hline
\multirow{16}{*}{0.07}	& \multirow{4}{*}{3$\times10^{-4}$}	& 2.0$\times10^{-4}$	& 2.61$\times10^{-1}$	& 0.414			& 0.349	& 3.16$\times10^{-2}$	& 0.345 \\
						& 									& 4.0$\times10^{-4}$	& 8.27$\times10^{-2}$	& 0.541			& 0.324	& 6.17$\times10^{-2}$	& 0.329 \\
						& 									& 6.0$\times10^{-4}$	& 3.39$\times10^{-2}$	& 0.645			& 0.330	& 8.00$\times10^{-2}$	& 0.326 \\
						& 									& 1.0$\times10^{-3}$	& 7.37$\times10^{-3}$	& 0.882			& 0.335	& 1.06$\times10^{-1}$	& 0.338 \\
\cline{2-8}
						& \multirow{3}{*}{6$\times10^{-4}$}	& 2.0$\times10^{-4}$	& 3.53$\times10^{-1}$	& 0.369			& 0.349	& 2.10$\times10^{-2}$	& 0.354 \\
						& 									& 5.0$\times10^{-4}$	& 8.81$\times10^{-2}$	& 0.568			& 0.325	& 5.65$\times10^{-2}$	& 0.330 \\
						& 									& 1.0$\times10^{-3}$	& 1.72$\times10^{-2}$	& 0.800			& 0.339	& 8.97$\times10^{-2}$	& 0.341 \\
\cline{2-8}
						& \multirow{9}{*}{1$\times10^{-3}$}	& 1.0$\times10^{-4}$	& 6.79$\times10^{-1}$	& \textendash\	& 0.386	& 4.21$\times10^{-3}$	& 0.386 \\
						& 									& 2.0$\times10^{-4}$	& 4.29$\times10^{-1}$	& 0.278			& 0.332	& 1.53$\times10^{-2}$	& 0.332 \\
						& 									& 3.0$\times10^{-4}$	& 2.71$\times10^{-1}$	& 0.408			& 0.324	& 2.80$\times10^{-2}$	& 0.324 \\
						& 									& 4.0$\times10^{-4}$	& 1.82$\times10^{-1}$	& 0.469			& 0.324	& 3.89$\times10^{-2}$	& 0.324 \\
						& 									& 6.0$\times10^{-4}$	& 9.22$\times10^{-2}$	& 0.563			& 0.329	& 5.50$\times10^{-2}$	& 0.335 \\
						& 									& 8.0$\times10^{-4}$	& 5.10$\times10^{-2}$	& 0.649			& 0.335	& 6.75$\times10^{-2}$	& 0.336 \\
						& 									& 1.0$\times10^{-3}$	& 2.96$\times10^{-2}$	& 0.728			& 0.339	& 7.83$\times10^{-2}$	& 0.343 \\
						& 									& 2.0$\times10^{-3}$	& 2.80$\times10^{-3}$	& 1.148			& 0.385	& 1.19$\times10^{-1}$	& 0.382 \\
						& 									& 3.0$\times10^{-3}$	& 1.36$\times10^{-3}$	& 1.649			& 0.451	& 1.39$\times10^{-1}$	& 0.455 \\
\hline
\multirow{10}{*}{0.10}	& \multirow{8}{*}{3$\times10^{-4}$}	& 3.0$\times10^{-4}$	& 4.74$\times10^{-1}$	& 0.244			& 0.541	& 2.19$\times10^{-2}$	& 0.546 \\
						& 									& 4.0$\times10^{-4}$	& 3.40$\times10^{-1}$	& 0.496			& 0.517	& 3.48$\times10^{-2}$	& 0.510 \\
						& 									& 5.0$\times10^{-4}$	& 2.49$\times10^{-1}$	& 0.583			& 0.493	& 4.75$\times10^{-2}$	& 0.496 \\
						& 									& 6.0$\times10^{-4}$	& 1.90$\times10^{-1}$	& 0.620			& 0.485	& 5.92$\times10^{-2}$	& 0.485 \\
						& 									& 8.0$\times10^{-4}$	& 1.24$\times10^{-1}$	& 0.675			& 0.469	& 7.74$\times10^{-2}$	& 0.470 \\
						& 									& 1.0$\times10^{-3}$	& 8.68$\times10^{-2}$	& 0.712			& 0.463	& 9.00$\times10^{-2}$	& 0.463 \\
						& 									& 2.0$\times10^{-3}$	& 1.82$\times10^{-2}$	& 0.865			& 0.432	& 1.35$\times10^{-1}$	& 0.433 \\
						& 									& 3.0$\times10^{-3}$	& 7.95$\times10^{-3}$	& 0.991			& 0.488	& 1.62$\times10^{-1}$	& 0.496 \\
\cline{2-8}
						& \multirow{2}{*}{6$\times10^{-4}$}	& 3.0$\times10^{-4}$	& 5.73$\times10^{-1}$	& \textendash\	& 0.560	& 1.35$\times10^{-2}$	& 0.565 \\
						& 									& 6.0$\times10^{-4}$	& 2.86$\times10^{-1}$	& 0.558			& 0.485	& 4.12$\times10^{-2}$	& 0.485 \\
						& 									& 1.0$\times10^{-3}$	& 1.42$\times10^{-1}$	& 0.676			& 0.474	& 7.15$\times10^{-2}$	& 0.470 \\
						& 									& 2.0$\times10^{-3}$	& 3.84$\times10^{-2}$	& 0.813			& 0.458	& 1.13$\times10^{-1}$	& 0.453 \\
\cline{2-8}
						& \multirow{10}{*}{1$\times10^{-3}$}	& 3.0$\times10^{-4}$	& 6.42$\times10^{-1}$	& \textendash\	& 0.591	& 9.33$\times10^{-3}$	& 0.583 \\
						& 									& 4.0$\times10^{-4}$	& 5.40$\times10^{-1}$	& \textendash\	& 0.529	& 1.56$\times10^{-2}$	& 0.528 \\
						& 									& 5.0$\times10^{-4}$	& 4.49$\times10^{-1}$	& 0.349			& 0.509	& 2.28$\times10^{-2}$	& 0.503 \\
						& 									& 6.0$\times10^{-4}$	& 3.74$\times10^{-1}$	& 0.470			& 0.491	& 3.02$\times10^{-2}$	& 0.496 \\				
						& 									& 8.0$\times10^{-4}$	& 2.68$\times10^{-1}$	& 0.565			& 0.478	& 4.46$\times10^{-2}$	& 0.478 \\
						& 									& 1.0$\times10^{-3}$	& 2.00$\times10^{-1}$	& 0.620			& 0.478	& 5.69$\times10^{-2}$	& 0.470 \\
						& 									& 1.5$\times10^{-3}$	& 1.08$\times10^{-1}$	& 0.707			& 0.466	& 8.03$\times10^{-2}$	& 0.472 \\
						& 									& 2.0$\times10^{-3}$	& 6.25$\times10^{-2}$	& 0.767			& 0.468	& 9.78$\times10^{-2}$	& 0.467 \\
						& 									& 2.5$\times10^{-3}$	& 3.85$\times10^{-2}$	& 0.826			& 0.471	& 1.13$\times10^{-1}$	& 0.476 \\
						& 									& 3.0$\times10^{-3}$	& 2.44$\times10^{-2}$	& 0.885			& 0.478	& 1.26$\times10^{-1}$	& 0.486 \\
\hline
\multirow{9}{*}{0.12}		& \multirow{4}{*}{6$\times10^{-4}$}	& 8.0$\times10^{-4}$	& 3.55$\times10^{-1}$	& 0.573			& 0.627	& 3.96$\times10^{-2}$	& 0.620 \\
						& 									& 1.0$\times10^{-3}$	& 2.68$\times10^{-1}$	& 0.676			& 0.603	& 5.39$\times10^{-2}$	& 0.595 \\
						& 									& 1.5$\times10^{-3}$	& 1.55$\times10^{-1}$	& 0.782			& 0.578	& 8.31$\times10^{-2}$	& 0.578 \\
						& 									& 2.0$\times10^{-3}$	& 1.05$\times10^{-1}$	& 0.841			& 0.568	& 1.01$\times10^{-1}$	& 0.568 \\
\cline{2-8}
						& \multirow{5}{*}{1$\times10^{-3}$}	& 8.0$\times10^{-4}$	& 4.52$\times10^{-1}$	& 0.381			& 0.633	& 2.85$\times10^{-2}$	& 0.632 \\
						& 									& 1.0$\times10^{-3}$	& 3.60$\times10^{-1}$	& 0.561			& 0.608	& 3.99$\times10^{-2}$	& 0.614 \\
						& 									& 1.5$\times10^{-3}$	& 2.20$\times10^{-1}$	& 0.720			& 0.577	& 6.59$\times10^{-2}$	& 0.582 \\
						& 									& 2.0$\times10^{-3}$	& 1.51$\times10^{-1}$	& 0.783			& 0.572	& 8.74$\times10^{-2}$	& 0.568 \\
						& 									& 3.0$\times10^{-3}$	& 7.49$\times10^{-2}$	& 0.878			& 0.559	& 1.12$\times10^{-1}$	& 0.564 \\
\enddata
\tablenotetext{a}{$\Delta_\Sigma$ can be calculated only when $\delta_\Sigma<0.5$.}
\end{deluxetable}

\onecolumngrid
\section{Temporal Changes of the Gap Properties \label{ap:time}}

\begin{figure}
	\centering \includegraphics[width=0.9\linewidth]{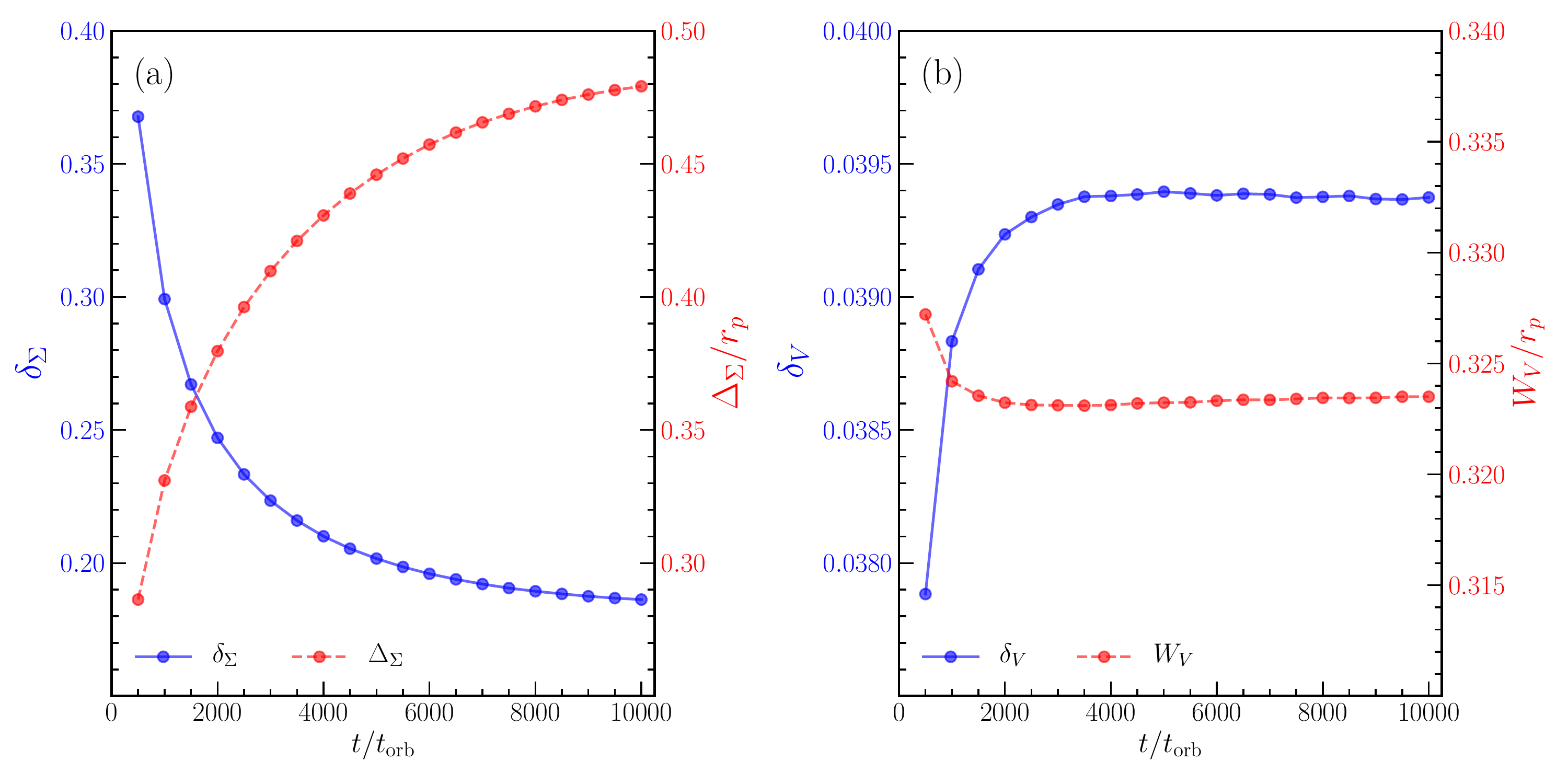}
	\caption{Temporal variations of (a) the depth $\delta_\Sigma$ and width $\Delta_\Sigma$ of the surface density gap and (b) the depth $\delta_V$ and width $W_V$ of the velocity gap in a model with $h_p/r_p=0.07$, $\alpha=1\times10^{-3}$ and $M_p=0.4\,M_J$. In each panel, the blue solid (left axis) and red dotted lines (right axis) draw the depth and width, respectively.}
	\label{fig:gaptime}
\end{figure}

To explore how the gap properties change with time, we select a model with $h_p/r_p=0.07$, $\alpha=1\times10^{-3}$ and $M_p=0.4\,M_J$ and measure the depth $\delta_\Sigma$ and width $\Delta_\Sigma$ of the surface density gap and the depth $\delta_V$ and width $W_V$ of the velocity gap at every $500\torb$ starting from $t=500\torb$ to $10^4\torb$. Figure \ref{fig:gaptime} plots the resulting temporal variations of (a) $\delta_\Sigma$ and $\Delta_\Sigma$ and (b) $\delta_V$ and $W_V$. The properties of the density gap converge to relatively slowly with time to quasi-steady values reached at around $t\sim8000\torb$. In this model, $\delta_\Sigma$ and $\Delta_\Sigma$ at $t=10^3\torb$ are about $\sim 1.6$ times larger and $\sim 0.7$ times smaller than the values at $t=10^4\torb$, respectively. Interestingly, the gap properties in the velocity profiles converge rapidly within $t=10^3\torb$: relative differences of $\delta_V$ and $W_V$ between $t=10^3$ and $10^4\torb$ are only 1.3\% and 0.22\%, respectively. In our simulations, the density profile deepens secularly with time, while retaining its radial gradient as well as the corresponding velocity profile intact.


\newpage

\begin{thebibliography}{}
\bibitem[ALMA Partnership et al.(2015)]{al2015} ALMA Partnership, Brogan, C. L., P\'{e}rez, L. M., et al. 2015, \apjl, 808, L3
\bibitem[Andrews et al.(2009)]{an2009} Andrews, S. M., Wilner, D. J., Hughes, A. M., Qi, C., Dullemond, C. P. 2009, \apj, 700, 1502
\bibitem[Andrews et al.(2010)]{an2010} Andrews, S. M., Wilner, D. J., Hughes, A. M., Qi, C., Dullemond, C. P. 2010, \apj, 723, 124
\bibitem[Andrews et al.(2012)]{an2012} Andrews, S. M., Wilner, D. J., Hughes, A. M., et al. 2012, \apj, 744, 162
\bibitem[Andrews et al.(2016)]{an2016} Andrews, S. M., Wilner, D. J., Zhu, Z., et al. 2016, \apjl, 820, L40
\bibitem[Bae et al.(2017)]{ba2017} Bae, J., Zhu, Z., \& Hartmann, L. 2017, \apj, 850, 201
\bibitem[Bae \& Zhu(2018)]{bz2018} Bae, J., \& Zhu, Z. 2018, \apj, 859, 118
\bibitem[Bae et al.(2018)]{ba2018} Bae, J., Pinilla, P., \& Birnstiel, T. 2018, \apjl, 864, L26
\bibitem[Ben\'{i}tez-Llambay \& Masset(2016)]{bm2016} Ben\'{i}tez-Llambay, P., \& Masset, F. 2016, \apjs, 223, 11
\bibitem[Bensity et al.(2018)]{be2018} Benisty, M., Juh\'{a}sz, A., Facchini, S., et al. 2018, \aap, 619, A171
\bibitem[Bergin et al.(2013)]{be2013} Bergin, E. A., Cleeves, L. I., Gorti, U., et al. 2013, \nat, 493, 644.
\bibitem[Casassus et al.(2013)]{ca2013} Casassus, S., van der Plas, G., S, P. M., et al. 2013, \nat, 493, 191
\bibitem[Cieza et al.(2017)]{ci2017} Cieza, L. A., Casassus, S., P\'{e}rez, S., et al. 2017, \apj, 851, L23
\bibitem[de Val-Borro et al.(2006)]{de2006} de Val-Borro, M., Edgar, R. G., Artymowicz, P., et al. 2006, \mnras, 370, 529
\bibitem[Dartois et al.(2003)]{da2003}Dartois, E., Dutrey, A., \& Guilloteau, S. 2003, \aap, 399, 773
\bibitem[Dong et al.(2011)]{do2011} Dong, R., Rafikov, R. R., \& Stone, J. M. 2011,\apj, 741, 57
\bibitem[Dong \& Fung(2017)]{df2017} Dong, R., \& Fung, J. 2017, \apj, 835, 146
\bibitem[Dong et al.(2017)]{do2017} Dong, R., Li, S., Chiang, E., \& Li, H. 2017, \apj, 843, 127
\bibitem[Dong et al.(2018a)]{do2018a} Dong, R., Liu, S.-y., Eisner, J., et al. 2018a, \apj, 860, 124
\bibitem[Dong et al.(2018b)]{do2018b} Dong, R., Li, S., Chiang, E., Li, H. 2018b, \apj, 866, 110
\bibitem[Duffell \& MacFadyen(2013)]{dm2013} Duffell, P. C., \& MacFadyen, A. I. 2013, \apj, 769, 41
\bibitem[Fedele et al.(2017)]{fe2017} Fedele, D., Carney, M., Hogerheijde, M. R., et al. 2017, \aap, 600, A72
\bibitem[Fedele et al.(2018)]{fe2018} Fedele, D., Tazzari, M., Booth, R., et al. 2018, \aap, 610, A24
\bibitem[Flaherty et al.(2017)]{fl2017} Flaherty, K. M., Hughes, A. M., Rose, S. C., et al. 2017, \apj, 843, 150
\bibitem[Flock et al.(2015)]{fl2015} Flock, M., Ruge, J. P., Dzyurkevich, N., et al. 2015, \aap, 574, A68
\bibitem[Fung et al.(2014)]{fu2014} Fung, J., Shi, J.-M., \& Chiang, E. 2014, \apj, 782, 88
\bibitem[Gaia Collaboration et al.(2018)]{gaia2018} Gaia Collaboration, Brown, A. G. A., Vallenari, A., et al. 2018, \aap, 616, A1
\bibitem[Ginski et al.(2016)]{gi2016} Ginski, C., Stolker, T., Pinilla, P., et al. 2016, \aap, 595, A112
\bibitem[Goldreich \& Tremaine(1980)]{gt1980} Goldreich, P., \& Tremaine, S. 1980, \apj, 241, 425
\bibitem[Gonzalez et al.(2015)]{go2015} Gonzalez, J.-F., Laibe, G., Maddison, S. T., Pinte, C., \& M\'{e}nard, F. 2015, \mnras, 454, L36
\bibitem[Goodman \& Rafikov(2001)]{gr2001} Goodman, J., \& Rafikov, R. R. 2001, \apj, 552, 793
\bibitem[Grady et al.(2013)]{gr2013} Grady, C. A., Muto, T., Hashimoto, J., et al. 2013, \apj, 762, 48
\bibitem[Huang et al.(2018)]{hu2018} Huang, J., Andrews, S. M., Dullemond, C. P., et al. 2018, \apjl, 869, L42
\bibitem[Isella et al.(2016)]{is2016} Isella, A., Guidi, G., Testi, L., et al. 2016, \prl, 117, 251101
\bibitem[Johansen et al.(2014)]{jo2014} Johansen, A., Blum, J., Tanaka, H., et al. 2014, in Protostars and Planets VI, ed. H. Beuther et al. (Tucson, AZ: Univ. Arizona Press), 547
\bibitem[Kanagawa et al.(2015a)]{ka2015a} Kanagawa, K. D., Tanaka, H., Muto, T., Tanigawa, T., \& Takeuchi, T. 2015b, \mnras, 448, 994
\bibitem[Kanagawa et al.(2015b)]{ka2015b} Kanagawa, K. D., Muto, T., Tanaka, H., et al. 2015a, \apjl, 806, L15
\bibitem[Kanagawa et al.(2016)]{ka2016} Kanagawa, K. D., Muto, T., Tanaka, H., et al. 2016, \pasj, 68, 43
\bibitem[Kanagawa et al.(2017)]{ka2017} Kanagawa, K. D., Tanaka, H., Muto, T., \& Tanigawa, T. 2017, \pasj, 69, 97
\bibitem[Kanagawa et al.(2018)]{ka2018} Kanagawa, K. D., Muto, T., Okuzumi, S., et al. 2018, \apj, 868, 48
\bibitem[Keppler et al.(2019)]{ke2019} Keppler, M., Teague, R., Bae., J., et al. 2019, \aap, 625, A118
\bibitem[Li et al.(2005)]{li2005} Li, H., Li, S., Koller, J., et al. 2005, \apj, 624, 1003
\bibitem[Lin \& Papaloizou(1979)]{lp1979} Lin, D. N. C., \& Papaloizou, J. 1979, \mnras, 188, 191
\bibitem[Lin \& Papaloizou(1986)]{lp1986} Lin, D. N. C., \& Papaloizou, J. 1986, \apj, 307, 395
\bibitem[Liu et al.(2018)]{liu2018} Liu, S.-F., Jin, S., Li, S., et al. 2018, \apj, 857, 87
\bibitem[Long et al.(2018)]{lo2018} Long, F., Pinilla, P., Herczeg, G. J., et al. 2018, \apj, 869, 17
\bibitem[Loomis et al.(2017)]{lo2017} Loomis, R. A., \"Oberg, K. I., Andrews, S. M., \& MacGregor, M. A. 2017, \apj, 840, 23
\bibitem[Lor\'{e}n-Aguilar \& Bate(2016)]{lb2016} Lor\'{e}n-Aguilar, P., \& Bate, M. R. 2016, \mnras, 457, L54
\bibitem[Lynden-Bell \& Pringle(1974)]{lp1974} Lynden-Bell, D., \& Pringle, J. E. 1974, \mnras, 168, 603
\bibitem[Marino et al.(2019)]{ma2019} Marino, S., Yelverton, B., Booth, M., et al. 2019, \mnras, 484, 1257
\bibitem[Masset (2000)]{m2000} Masset, F. 2000, \aaps, 141, 165
\bibitem[Miranda \& Rafikov(2019)]{mr2019} Miranda, R., \& Rafikov, R. R. 2019, \apj, 515, 767
\bibitem[Miotello et al.(2017)]{mi2017} Miotello, A., van Dishoeck, E. F., Williams, J. P., et al. 2017, \aap, 599, A113.
\bibitem[Muto et al.(2012)]{mu2012} Muto, T., Grady, C. A., Hashimoto, J., et al. 2012, \apjl, 748, L22
\bibitem[Nazari et al.(2019)]{na2019} Nazari, P., Booth., R. A., Clarke, C. J., et al. 2019, \mnras, 485, 5914
\bibitem[Nelson \& Papaloizou(2003)]{np2003} Nelson, R. P., \& Papaloizou, J. C. B. 2003, \mnras, 339, 993
\bibitem[Okuzumi et al.(2016)]{ok2016} Okuzumi, S., Momose, M., Sirono, S.-i., Kobayashi, H., \& Tanaka, H. 2016, \apj, 821, 82
\bibitem[Papaloizou \& Lin(1984)]{pap84} Papaloizou, J., \& Lin, D. N. C. 1984, \apj, 285, 818
\bibitem[P\'{e}rez et al.(2015)]{pe2015} P\'{e}rez, S., Dunhill, A., Casassus, S., et al. 2015, ApJ, 811, L5
\bibitem[P\'{e}rez et al.(2018)]{pe2018} P\'{e}rez, S., Casassus, S., \& Ben\'{ı}tez-Llambay, P. 2018, \mnras, 480, L12
\bibitem[Pinte et al.(2018)]{pi2018} Pinte, C., Price, D. J., Ménard, F., et al. 2018, \apj, 860, L13
\bibitem[Rafikov(2002)]{ra2002} Rafikov, R. R. 2002, \apj, 572, 566
\bibitem[Rosotti et al.(2016)]{ro2016} Rosotti, G. P., Juhasz, A., Booth, R. A., \& Clarke, C. J.\ 2016, \mnras, 459, 2790
\bibitem[Shakura \& Sunayev (1973)]{ss1973} Shakura, N. I., \& Sunyaev, R. A. 1973, \aap, 24, 337
\bibitem[Suriano et al.(2017)]{su2017} Suriano, S. S., Li, Z.-Y., Krasnopolsky, R., \& Shang, H. 2017, \mnras, 468, 3850
\bibitem[Sheehan \& Eisner(2018)]{se2018} Sheehan, P. D., \& Eisner, J. A. 2018, \apj, 857, 18
\bibitem[Takahashi \& Inutsuka(2016)]{ti2016} Takahashi, S. Z., \& Inutsuka, S.-i. 2016, \aj, 152, 184
\bibitem[Teague et al.(2018)]{te2018} Teague, R., Bae, J., Bergin, E., Birnstiel, T.,\& Foreman-Mackey, D. 2018, \apj, 860, L12
\bibitem[Uribe et al.(2011)]{ur2011} Uribe, A. L., Klahr, H., Flock, M., \& Henning, T. 2011, \apj, 736, 85
\bibitem[van den Ancker et al.(1997)]{van1997} van den Ancker M. E., The, P. S., Tjin A Djie, H. R. E., et al., 1995, \aap, 324, L33
\bibitem[van der Marel et al.(2013)]{van2013} van der Marel, N., van Dishoeck, E. F., Bruderer, S. et al. 2013, Science, 340,1199
\bibitem[van der Marel et al.(2019)]{van2019} van der Marel, N., Dong, R., di Francesco, J., et al. 2019, \apj, 872, 112
\bibitem[van Terwisga et al.(2018)]{va2018} van Terwisga, S. E., van Dishoeck, E. F., Ansdell, M., et al. 2018, \aap, 616, A88
\bibitem[Wagner et al.(2015)]{wa2015} Wagner, K., Apai, D., Kasper, M., \& Robberto, M. 2015, \apjl, 813, L2
\bibitem[Winters et  al.(2003)]{wi2003} Winters, W. F., Balbus, S. A.,\& Hawley, J. F. 2003, \apj, 589, 543
\bibitem[Youdin \& Goodman(2005)]{yg2005} Youdin, A. N., \& Goodman, J. 2005, \apj, 620, 459,
\bibitem[Zhang et al.(2015)]{zh2015} Zhang, K., Blake, G. A., \& Bergin, E. A. 2015, \apjl, 806, L7
\bibitem[Zhang et al.(2018)]{zha2018} Zhang, S., Zhu, Z., Huang, J. et al. 2018, \apjl, 869, L47
\bibitem[Zhu et al.(2013)]{zh2013} Zhu, Z., Stone, J. M.,\& Rafikov, R. R. 2013, \apj, 768, 143
\bibitem[Zhu(2018)]{zh2018} Zhu, Z. 2018, \mnras, 483, 4221
\end{thebibliography}
\end{document}